\documentclass[twocolumn,aps,superscriptaddress,floatfix]{revtex4-1}
\usepackage{array}
\usepackage{amsmath}
\usepackage{amssymb}
\usepackage{graphicx}
\usepackage{color}
\usepackage{adjustbox}
\usepackage{lipsum}

\newcommand{\be}{\begin{equation}}
\newcommand{\ee}{\end{equation}}
\newcommand{\bea}{\begin{eqnarray}}
\newcommand{\eea}{\end{eqnarray}}
\newcommand{\bse}{\begin{subequations}}
\newcommand{\ese}{\end{subequations}}

\newcommand{\emb}{EuMg$_2$Bi$_2$}
\newcommand{\esa}{EuSn$_2$As$_2$}
\newcommand{\nsa}{NaSn$_2$As$_2$}

\begin{document}

\title{A-type antiferromagnetic order and magnetic phase diagram of the trigonal Eu spin-7/2 triangular-lattice compound  EuSn$_2$As$_2$}

\author{Santanu Pakhira}
\affiliation{Ames Laboratory, Iowa State University, Ames, Iowa 50011, USA}
\author{M. A. Tanatar}
\affiliation{Ames Laboratory, Iowa State University, Ames, Iowa 50011, USA}
\affiliation{Department of Physics and Astronomy, Iowa State University, Ames, Iowa 50011, USA}
\author{Thomas Heitmann}
\affiliation{The Missouri Research Reactor and Department of Physics and Astronomy, University of Missouri, Columbia, Missouri 65211, USA}
\author{David Vaknin}
\affiliation{Ames Laboratory, Iowa State University, Ames, Iowa 50011, USA}
\affiliation{Department of Physics and Astronomy, Iowa State University, Ames, Iowa 50011, USA}
\author{D. C. Johnston}
\affiliation{Ames Laboratory, Iowa State University, Ames, Iowa 50011, USA}
\affiliation{Department of Physics and Astronomy, Iowa State University, Ames, Iowa 50011, USA}

\date{\today}

\begin{abstract}

The trigonal compound \esa\ was recently discovered to host Dirac surface states within the bulk band gap and orders antiferromagnetically below the N\'eel temperature $T_{\rm N} \approx 24$~K\@. Here the magnetic ground state of single-crystal \esa\ and the evolution of its properties versus temperature~$T$ and applied magnetic field~$H$ are reported.  Included are zero-field single-crystal neutron diffraction measurements versus~$T$,  magnetization $M(H,T)$, magnetic susceptibility $\chi(H,T) = M(H,T)/H$, heat capacity $C_{\rm p}(H, T)$, and electrical resistivity $\rho (H, T)$ measurements. The neutron-diffraction and $\chi(T)$ measurements both indicate a collinear A-type antiferromagnetic (AFM) structure below $T_{\rm N}=23.5(2)$~K, where the Eu$^{2+}$ spins~$S=7/2$ in a triangular $ab$-plane layer (hexagonal unit cell) are aligned ferromagnetically in the $ab$~plane whereas the spins in adjacent Eu planes along the $c$~axis are aligned antiferromagnetically.  The $\chi(H_{ab},T)$ and $\chi(H_c,T)$ data together indicate a smooth crossover between the collinear AFM alignment and an unknown magnetic structure at $H \sim 0.15$~T\@.  Dynamic spin fluctuations up to 60~K are evident in the $\chi(T)$, $C_{\rm p}(T)$ and $\rho(H,T)$ measurements, a temperature that is more than twice $T_{\rm N}$\@. The  $\rho (H, T)$ of the compound does not reflect a contribution of the topological state, but rather is consistent with a low-carrier-density metal with strong magnetic scattering. The magnetic phase diagrams  for both $H\parallel c$ and $H \parallel ab$  in the $H$-$T$ plane are constructed from the $T_{\rm N}(H)$, $\chi(H,T)$, $C_{\rm p}(H, T)$, and $\rho(H,T)$ data.

\end{abstract}

\maketitle

\section{Introduction}

Rare-earth intermetallic compounds are of considerable interest due to the observations of complex magnetism, superconductivity, heavy-fermion behavior, Kondo effect, valence fluctuation, large magnetocaloric effect, magnetoresistance, and quantum criticality~\cite{Fertig1977_IC, Nagarajan1994_IC, Ghosh1995_IC, Curro2000_IC, Hundley2001_IC, Sakai2011_IC, Yamaoka2014_IC, Pecharsky1997_IC, Pakhira2016_IC, Pakhira2017_IC, Buschow1969_IC, Gignoux1984_IC, Ishida2002_IC, Arndt2011_IC}.   Besides $4f$-element magnetism, many rare-earth-based intermetallics have expanded in a new direction by hosting novel electronic states through a complex interplay of magnetism and band topology~\cite{Hirschberger2016_GdPtBi, Shekhar2018_RPtBi, Wang2016_YbMnBi2, May2014_EuMnBi2, Masuda2016_EuMnBi2, Soh2019_EuMnSb2, Schellenberg2011_EuCd2As2, Jo2020_EuCd2As2, Xu2019_EuIn2As2, Pakhira2020_EuMg2Bi2, Pakhira2021_EuMg2Bi2}. This interplay has generated immense interest due to the discovery of nontrivial quantum phenomena such as anomalous quantum Hall effect (AQHE), axion electrodynamics, and realization of relativistic particles like Majorana fermions~\cite{Hasan2010_TI, Fu2007_TI, Tokura2019_TI, Qi2006_QAHE, Liu2008_QAHE, Yu2010_QAHE, Essin2009_AED, Li2010_AED, Fu2008_Majorana,Akhmerov2009_Majorana, Cook2011_Majorana}. Topological states like Dirac surface points are expected to be gapped or split in the presence of magnetic order due to time-reversal-symmetry breaking. Thus studies of the nature of magnetic order and the associated phase diagram in magnetic topological compounds is essential to reveal the presence or absence of nontrivial electronic states.

\esa\ is an intrinsic rare-earth-based topological insulator (TI) that has been recently reported to exhibit Dirac surface states with energies within the bulk band gap~\cite{Li2019_EuSn2As2}. However, conventional Hall-effect measurements reveal the presence of hole carriers with a concentration $p=3.6\times10^{20}$~cm$^{-3}$ at 2~K, rendering the compound metallic up to at least 300~K with an in-plane resistivity of order 0.4~m$\Omega$\,cm~\cite{Chen2020_EuSn2As2}.  The hole carriers might arise, for example, from low concentrations of Sn vacancies.

\esa\ crystallizes in the layered \nsa-type crystal structure with trigonal space group $R\overline{3}m$\@. In this crystal structure (Fig.~\ref{Fig_crystal_structure}), the Eu atoms form a triangular lattice in the $ab$ plane (hexagonal unit cell) with each Eu atom coordinated by six As atoms  to form a trigonal prism. The Sn and As atoms are covalently bonded in a honeycomb arrangement and the Sn atoms between two adjacent Sn-As  layers face each other with inversion symmetry.  In such a lattice geometry, magnetic frustration can arise when the nearest-neighbor exchange interaction between the Eu ions is antiferromagnetic~(AFM). The compound is indeed reported to order antiferromagnetically below its N\'eel temperature  $T_{\rm N} \approx 24$~K~\cite{Li2019_EuSn2As2, Arguilla2017_EuSn2As2, Chen2020_EuSn2As2}. However, the nature of the AFM ordering and the phase diagram in the magnetic field $H$-temperature $T$ plane is not yet known. Interestingly, the magnetic susceptibility $\chi(T)$ below $T_{\rm N}$ is found to be quite different in three earlier reports where single crystals were grown in different ways. Comparing the $\chi(T)$ data reported in Refs.~\cite{Li2019_EuSn2As2, Arguilla2017_EuSn2As2, Chen2020_EuSn2As2}, the $ab$-plane $\chi_{ab}(T)$ shows a clear Curie-like upturn at $T < 10$~K and its magnitude depends on the single-crystal growth procedure. In order to understand the band topology and their magnetic field evolution in a magnetic TI, an understanding of the nature of magnetic ordering and the magnetic $H$-$T$ phase diagram is required. 

Herein we report room-temperature x-ray measurements of the crystallography, zero applied magnetic field~$H$ neutron-diffraction measurements of the magnetic structure at $T \lesssim T_{\rm N}$, and measurements of the magnetization $M(H,T)$, magnetic susceptibility \mbox{$\chi(H,T)=M(H,T)/H$,}  heat capacity $C_{\rm p}(H,T)$, and electrical resistivity $\rho(H,T)$. We find that \esa\ exhibits A-type AFM ordering below $T_{\rm N} = 23.5(2)$~K with the moments aligned ferromagnetically in the $ab$~plane with the spins in adjacent planes along the $c$-axis aligned antiferromagnetically. Depending on the crystal-growth procedure, different amounts of low-temperature Curie-Weiss-like upturns in the low-field $\chi(T)$ are obtained, confirming previous results. Low-field $\chi(T<T_{\rm N})$ measurements suggest the presence of three-fold colliinear AFM domains.  In magnetic fields applied in the $ab$~plane the spins reorient or change the magnetic structure under a small field of $H = 0.12$~T\@. In addition to the drop in $\rho(T)$ on cooling below $T_{\rm N}$, weak low-field transitions with a very small entropy change were detected in the $\rho(H,T)$ data. The magnetic phase diagrams in the $H$-$T$ plane associated with AFM-PM phase boundary constructed from the observed magnetic, heat capacity and resistivity data are presented for both $H\parallel ab$ and $H\parallel c$.

The experimental details are presented in Sec.~\ref{Sec:ExpDet} and the structural and elemental composition characterization of our crystals in Sec.~\ref{Sec:Struct}.  The single-crystal neutron diffraction measurements of the magnetic structure are presented in Sec.~\ref{Sec:neutrons}, the $\chi(H,T)$ and $M(H,T)$ measurements in Sec.~\ref{Sec:chiM}, the $C_{\rm p}(H,T)$ measurements in Sec.~\ref{Cp}, and the $\rho(H,T)$ measurements in Sec.~\ref{rho}.  A summary and the magnetic $H\parallel c$-$T$ and $H\parallel ab$ phase diagrams constructed from the various measurements are presented in Sec.~\ref{Summary}.

\section{\label{Sec:ExpDet} Experimental details}

\esa\ single crystals were grown in three different ways using high-purity elemental Eu (Ames Laboratory, 99.99\%), Sn (Alfa Aesar, 99.999\%) , and As (Alfa Aesar, 99.9999\%). One batch of crystals, labeled as ``Crystal A'' were grown with a starting ratio of Eu:Sn:As = 1:3:20, as utilized previously~\cite{Li2019_EuSn2As2}.  In this process, the starting elements were loaded into an alumina crucible which was sealed in a silica tube under $\approx$ 1/4 atm of Ar gas. The sealed tube was then heated to 1000~$^{\circ}$C at a rate 100~$^{\circ}$C/h where the temperature was held for 12 h, followed by cooling to 750~$^{\circ}$C/h at a cooling rate of 2~$^{\circ}$C/h. The silica tube containing the crucible was then centrifuged to remove the crystals from the remaining flux. Clean shiny crystals of typical dimensions $2\times1\times0.5$~mm$^3$ were obtained by this method.

Another batch of crystals denoted ``Crystal B'' was grown with a molar ratio Eu:Sn:As = 1.1:2:20 where the silica tube containing the loaded alumina crucible was initially heated to 600~$^{\circ}$C at a rate 50~$^{\circ}$C/h, held there for 24~h, and then heated to 1050~$^{\circ}$C at a rate 50~$^{\circ}$C/h followed by a dwell of 48 h at that temperature. Finally the tube was cooled to 600~$^{\circ}$C at a rate of 2~$^{\circ}$C/h and single crystals of typical size $1\times1 \times 0.5$~mm$^3$ were obtained after centrifuging the excess flux.

A third batch of  \esa\ single crystals denoted ``Crystal C'' were synthesized with a stoichiometric starting ratio of the elements (Eu:Sn:As = 1:2:2) in an alumina crucible that was sealed under $\approx 1/4$~atm of Ar in a silica tube~\cite{Arguilla2017_EuSn2As2}. In this growth, a sealed silica tube containing the filled crucible was heated to 850~$^{\circ}$C at a rate 50~$^{\circ}$C/h and the temperature held there for 12~h. Then the silica-tube assembly was slowly cooled to room temperature in 100~h. The crystals were separated by mechanical cleaving. A few thin single crystals were selected for measurements with dimensions of $3 \times 2 \times 0.2$~mm$^3$.

We grew \esa\ crystals using the three methods described above in an attempt to obtain the smallest paramagnetic Curie-like term at low~$T$ which we found to be the crystals~C\@.

Room-temperature powder x-ray diffraction (XRD) measurements were carried out on ground crystals using a Rigaku Geigerflex x-ray diffractometer with Cu-$K_\alpha$ radiation. Crystal-structure analysis was performed by Rietveld refinement using the FULLPROF software package~\cite{Carvajal1993}.  The chemical composition and homogeneity of the crystals were measured with a JEOL scanning electron microscope (SEM) equipped with an EDS (energy-dispersive x-ray spectroscopy) analyzer. A Magnetic-Properties-Measurement System (MPMS) from  Quantum Design, Inc., was used for $\chi(T)$ and $M(H)$ measurements in the $T$ range 1.8--300 K and with $H$ up to 5.5 T (1~T~$\equiv10^4$~Oe). A physical properties measurement system (PPMS, Quantum Design, Inc.) was used to measure $C_{\rm p}(H,T)$ and $\rho(H,T)$ in the $T$ range 1.8--300~K and in the $H$ range 0--9~T\@. The  $C_{\rm p}(H,T)$ was measured by a relaxation technique using the PPMS\@.

Samples for electrical resistivity measurements were cut into rectangular parallelepiped shapes with typical dimensions 2$\times$0.5$\times$0.1 mm$^3$. The long axis was parallel to the $ab$~plane and the short axis was along the $c$~axis. Contacts to the samples were made using silver wires soldered to the sample with tin \cite{Tanataranisotropy,SUST}. For measurements in magnetic fields in the $H \bot c$ direction, samples were attached to a plastic support using Apiezon N~grease. Four-probe resistivity measurements were performed using a Quantum Design, Inc., Physical Property Measurement System (PPMS).

Single-crystal neutron-diffraction experiments were performed in zero applied magnetic field using the TRIAX triple-axis spectrometer at the University of Missouri Research Reactor (MURR). An incident neutron beam of energy $E_i = 30.5$ meV was directed at the sample using a pyrolytic graphite (PG) monochromator.  A pyrolytic-graphite (PG) analyzer was used to reduce the background. Neutron wavelength harmonics were removed from the beam using PG filters placed before the monochromator and between the sample and analyzer. Beam divergence was limited using collimators before the monochromator; between the monochromator and sample; sample and analyzer; and analyzer and detector of $60^\prime-60^\prime-40^\prime-40^\prime$, respectively.

A 20~mg \esa\ crystal (Crystal C) was mounted on the cold tip of an Advanced Research Systems closed-cycle refrigerator with a base temperature of 4~K\@. The crystal was aligned in the $(HH0)$ and $(00L)$ planes. Due to strong absorption by Eu, we could only identify a few peaks and the alignment relied on the (1~1~0) and (0~0~15) Bragg peaks using the room-temperature \esa\ lattice parameters $a = 4.207(3)$~\AA\ and \mbox{$c = 26.463(5)$~\AA} and the base-temperature lattice parameters $a = 4.178(3)$~\AA\ and $c = 26.233(5)$~\AA.

\section{\label{Sec:Struct} Structural and elemental characterization}

\begin{figure}
\includegraphics[width = 3.4in]{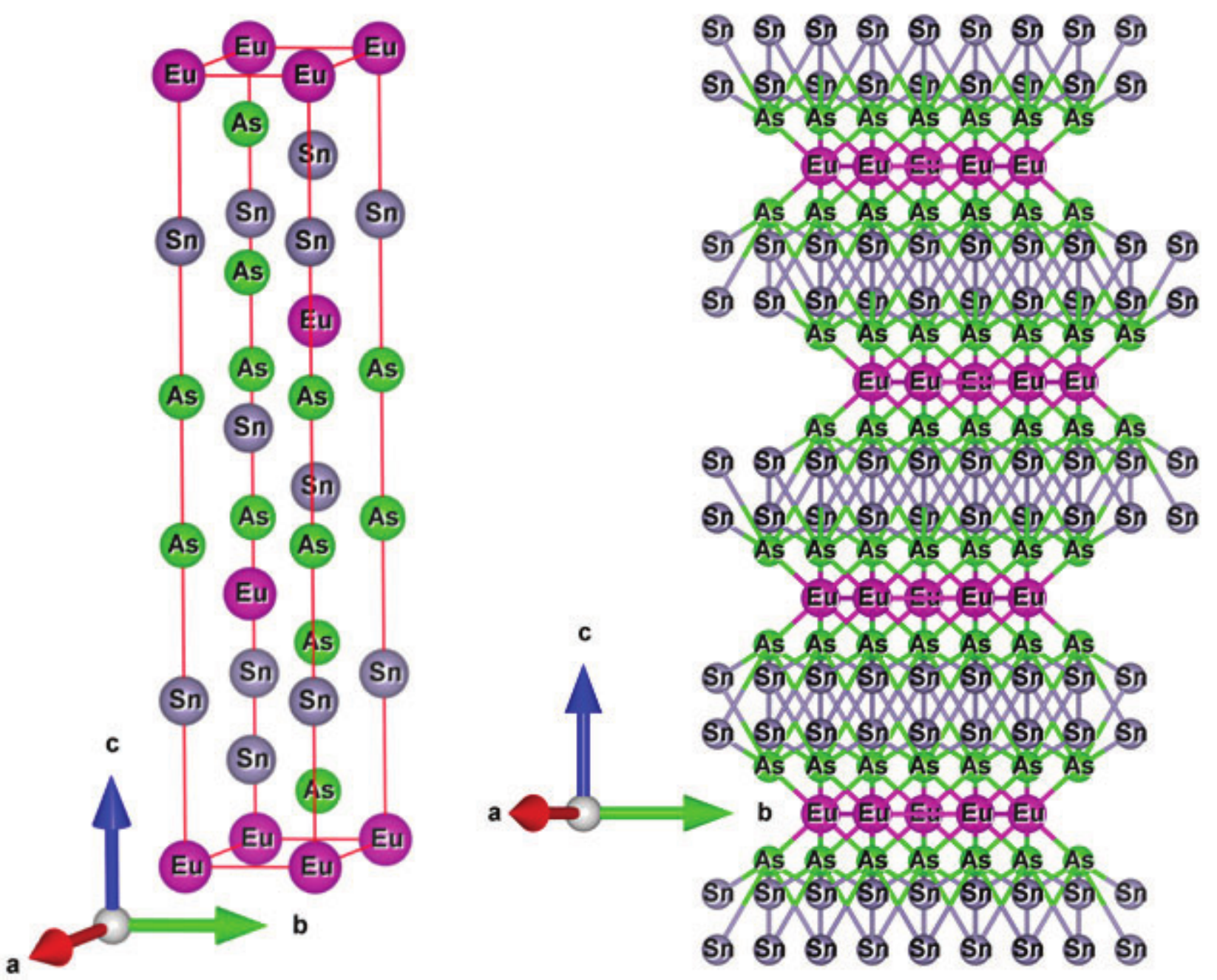}
\caption{Unit cell and crystal structure of \esa. Eu atoms form a hexagonal layer situated between two buckled SnAs honeycomb layers. The Eu atoms are colored dark purple, the Sn atoms blue, and the As atoms green.  The crystal structure was drawn using the VESTA program~\cite{Momma2011}.}
\label{Fig_crystal_structure}
\end{figure}

\begin{figure}
\includegraphics[width = 0.45\textwidth]{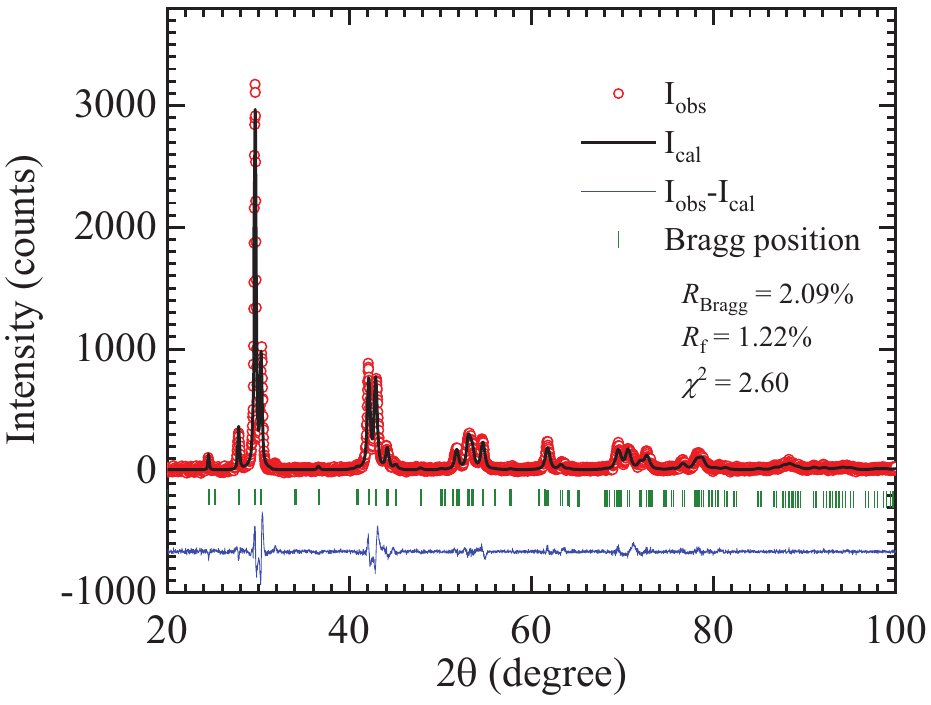}
\caption{Room-temperature XRD pattern of powdered \esa\ Crystals~C along with the refinement of the data using the FULLPROF software. The red circles are the experimental data points, the black solid line is the calculated refinement based on the \nsa\ crystal structure (space group $R\overline{3}m$), the blue solid curve is the difference between the experimental and calculated diffraction patterns, and the green vertical bars represent the allowed Bragg positions.  The fit-quality parameters are given in the figure.}
\label{Fig_XRD}
\end{figure}

The crystal structure of \esa\ is shown in Fig.~\ref{Fig_crystal_structure}~\cite{Momma2011}. The room-temperature powder XRD pattern taken on ground \esa\ single crystals (Crystals C) is shown in Fig.~\ref{Fig_XRD}. The experimental data were fitted using the FULLPROF software package based on the symmetry of the \nsa\--type crystal structure~\cite{Asbrand1995} having a trigonal lattice with space group $R\overline{3}m$.  The best fit does not reveal any significant impurity contribution in the studied crystals. The refined lattice parameters $a = b = 4.2006(3)$\AA~ and $c = 26.394(2)$~\AA\ are in good agreement with reported values~\cite{Arguilla2017_EuSn2As2}. The SEM-EDS scans further confirm the homogeneity of the crystals with average compositions EuSn$_{1.95(4)}$As$_{2.00(4)}$ for Crystal A, EuSn$_{1.96(5)}$As$_{2.00(5)}$ for Crystal B, and EuSn$_{1.95(5)}$As$_{2.02(5)}$ for Crystal C\@.

\section{\label{Sec:neutrons} Single-crystal neutron-diffraction measurements}

\begin{figure}
\centering
\includegraphics[width = 2.75 in]{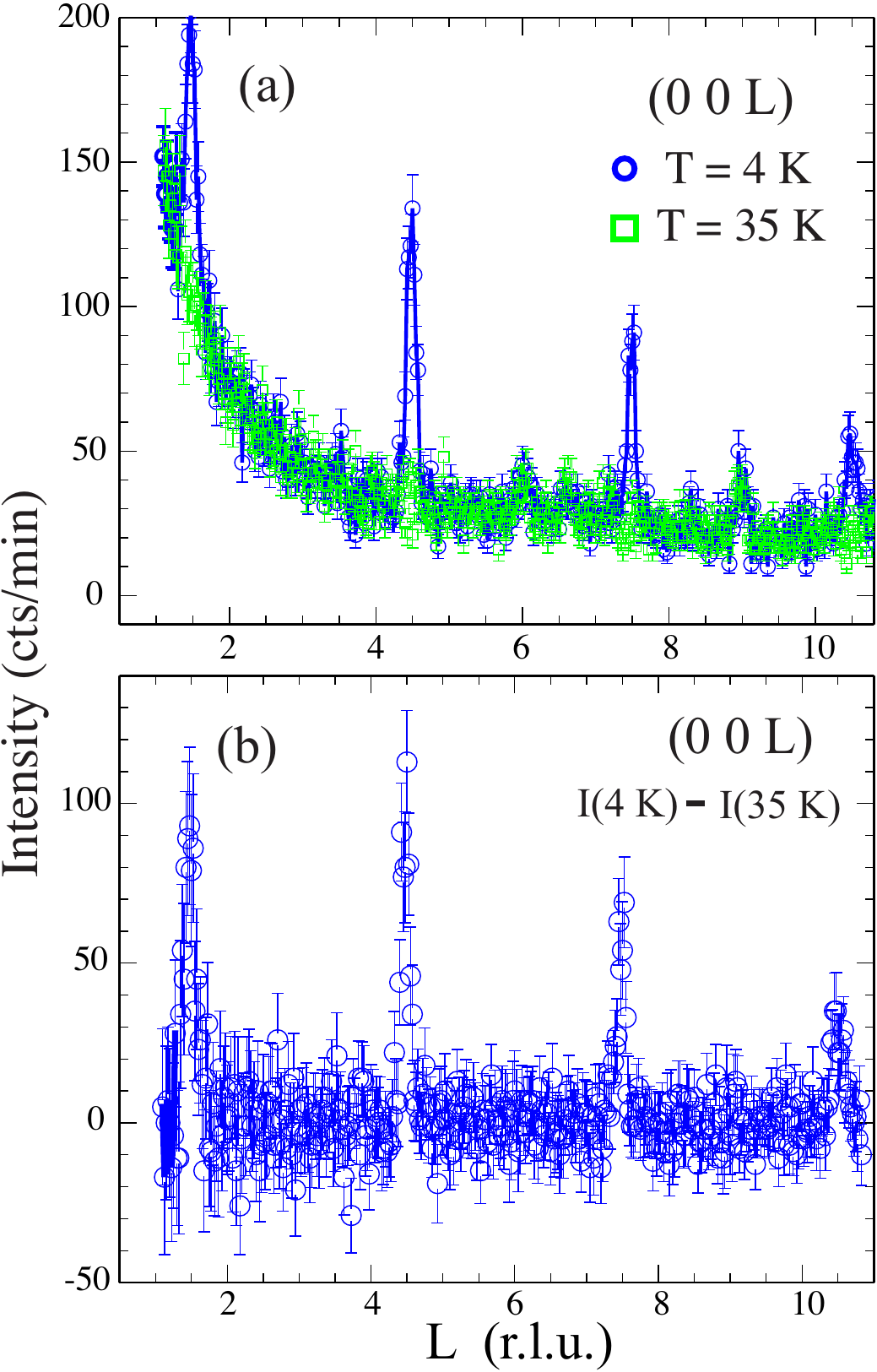}
\caption{(a) Diffraction patterns along $(0~0~L)$ reciprocal-lattice units of single-crystal \esa\ (Crystal C) at 4 and 36~K as indicated.  (b) Difference between the $(0~0~L)$ patterns at 4~K and 36~K only showing Bragg-diffraction reflections at $L = 1.5 + 3n \  (n = 1,2,3,\ldots)$.  }
\label{Fig:00l}
\end{figure}

\begin{figure}
\centering
\includegraphics[width = 2.75in]{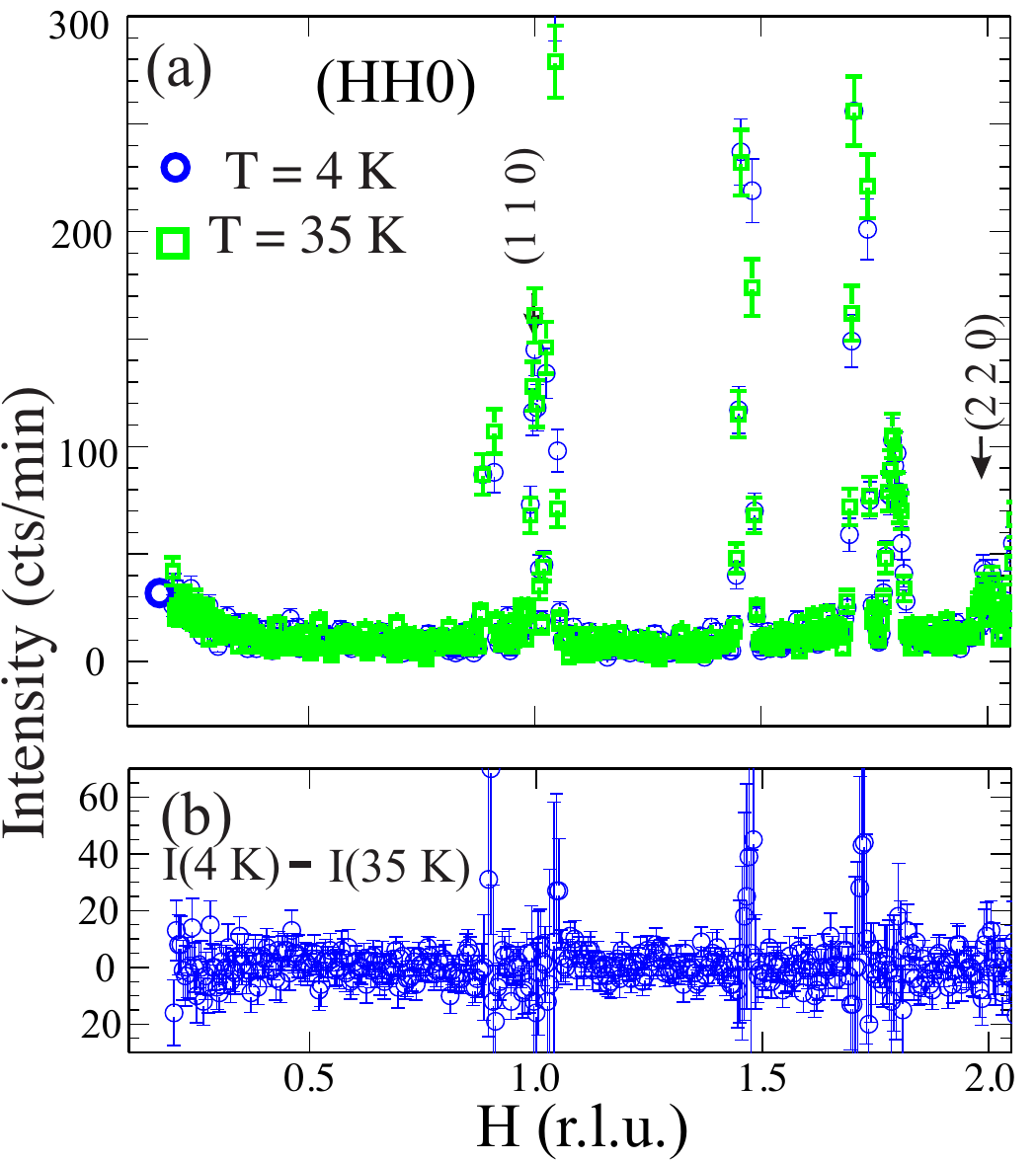}
\caption{(a)~Diffraction patterns along $(H~H~0)$ of single-crystal \esa\ at 4~K and 36~K  as indicated.  The extra peaks are from the aluminum sample holder. (b)~The difference between the $(H~H~0)$ patterns taken at 4~K and 35~K\@. The absence of emergent magnetic Bragg reflections at low temperature indicates that $ab$-plane layers of Eu moments are aligned ferromagnetically in the $ab$~plane and are stacked antiferromagnetically along the $c$~axis.}
\label{Fig:hh0}
\end{figure}

Figure~\ref{Fig:00l}(a) shows diffraction scans of Crystal C along $(0~0~L)$~r.l.u.\ (reciprocal-lattice units)  at 4 and 35~K, where reflections at  $L = 1.5 + 3n\ (n =1,2,3, \ldots)$ emerge at $T = 4$~K\@.  For more clarity, Fig.~\ref{Fig:00l}(b) shows the difference between these two scans, where within experimental uncertainty there is no evidence for other reflections associated with a modulated structure along the $c$~axis.  A similar scan at 4 and 35~K with expected magnetic reflections along  $(1~1~L)$ (not shown) shows very weak new signals at the magnetic Bragg reflections due to the decreasing form factor of Eu$^{2+}$ at larger momentum transfers~$Q$.

\begin{figure}
\centering
\includegraphics[width = 3.4in]{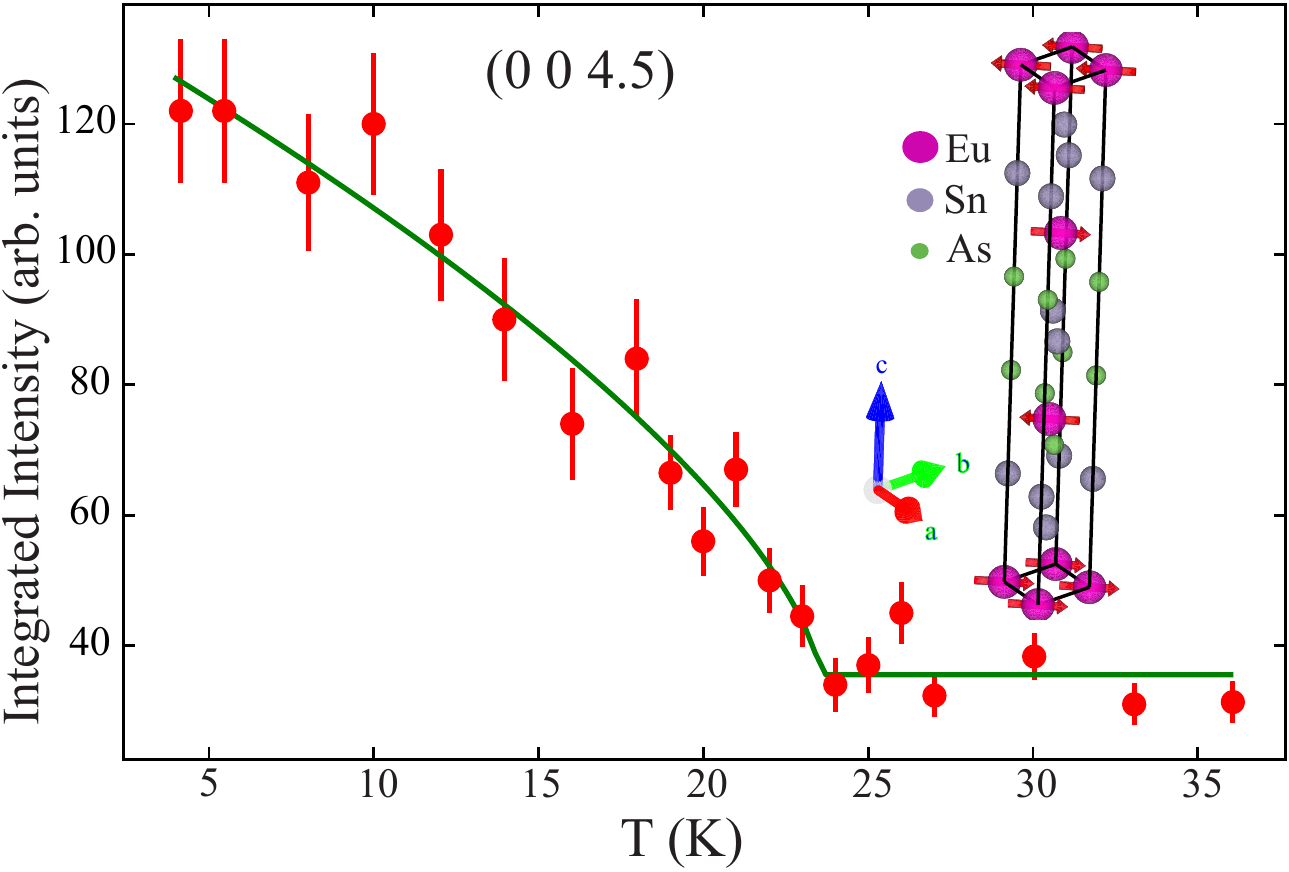}
\caption{Integrated intensity as a function of temperature $T$ of the (0 0 4.5)~rlu magnetic Bragg reflection. The solid curve is a phenomenological power-law fit by \mbox{$I(T) = C|1-T/T_{\rm N}|^{2\beta}$}, yielding  $T_{\rm N} = (23.5 \pm 0.2)$~K\@ and $\beta =0.33(1)$. The magnetic structure of the Eu$^{2+}$ ordered moments depicts half of the AFM unit cell. The neutron-diffraction data are not sufficient for determining the direction of the ferromagnetically-aligned moment in each $ab$~plane.}
\label{Fig:OP}
\end{figure}

Qualitatively, these newly emerging Bragg reflections indicate the doubling of the unit cell along the $c$~axis, and establish  that these Bragg reflections are associated with A-type AFM ordering with  propagation vector $\vec{\tau} = \left(0 0 \frac{1}{2}\right)$~r.l.u. The magnetic structure consists of $ab$-plane layers of Eu spins with the moments ferromagnetically aligned in the $ab$~plane that are stacked antiferromagnetically along the $c$~axis.  The in-plane direction of the FM moments cannot be determined from neutron diffraction. This AFM structure is equivalent to a $c$-axis helix with a turn angle between FM-aligned layers of $180^\circ$.   Our model calculations of the magnetic structure, shown in the inset of Fig.~\ref{Fig:OP}, predicts that whole and half-integer $(0~0~L)$ reflections are allowed, but are too weak to be detected with our small crystal. The calculations show that the most prominent reflections are the $L = 1.5 + 3n\ (n =1,2,3, \ldots)$ as observed experimentally.  Using this model we estimate the average ordered magnetic moment at $T=4$~K to be $g\langle S \rangle = (6.7 \pm 0.5)~\mu_{\rm B}$, where $g$ is the spectroscopic splitting factor of the Eu spin and $\langle S\rangle$ is the expectation value of the spin angular momentum in units of $\hbar$.  This value agrees within the error with the value of 7~$\mu_{\rm B}$ obtained using $g=2$ and $S=7/2$.  We note that our modeling precludes a helical structure (in zero magnetic field) assuming that each of the six FM layers is rotated by 60 degrees with respect to the previous layer, as such a structure predicts other prominent half integer $(0~0~L)$ peaks that are not observed experimentally.

Figure~\ref{Fig:hh0} shows that no additional magnetic peaks are observed in the difference between scans taken at 4 and 35~K along $(H~H~0)$, consistent with a single AFM propagation vector~$\vec{\tau}$. Other scans along $(\mbox{0.5~0.5~}~L)$ and \mbox{($H$~$H$~4.5)} (not shown) are consistent with a single $\vec{\tau}$.  Figure~\ref{Fig:OP} shows the integrated intensity versus~$T$ of the (0~0~4.5) magnetic Bragg reflection. The solid curve is a power-law fit $I(T) = C(1-T/T_{\rm N})^{2\beta}$ yielding $T_{\rm N} = 23.5(2)$~K and $\beta =0.33(1)$, where the latter value is consistent with three-dimensional ordering of Heisenberg spins.  The magnetic structure depicted in Figure~\ref{Fig:OP} corresponds to half the magnetic unit cell.

\section{\label{Sec:chiM} Magnetic Susceptibility and Magnetization Measurements}

\subsection{Magnetic susceptibility in the high-temperature paramagnetic state}

\begin{figure}
\includegraphics[width = 3.3in]{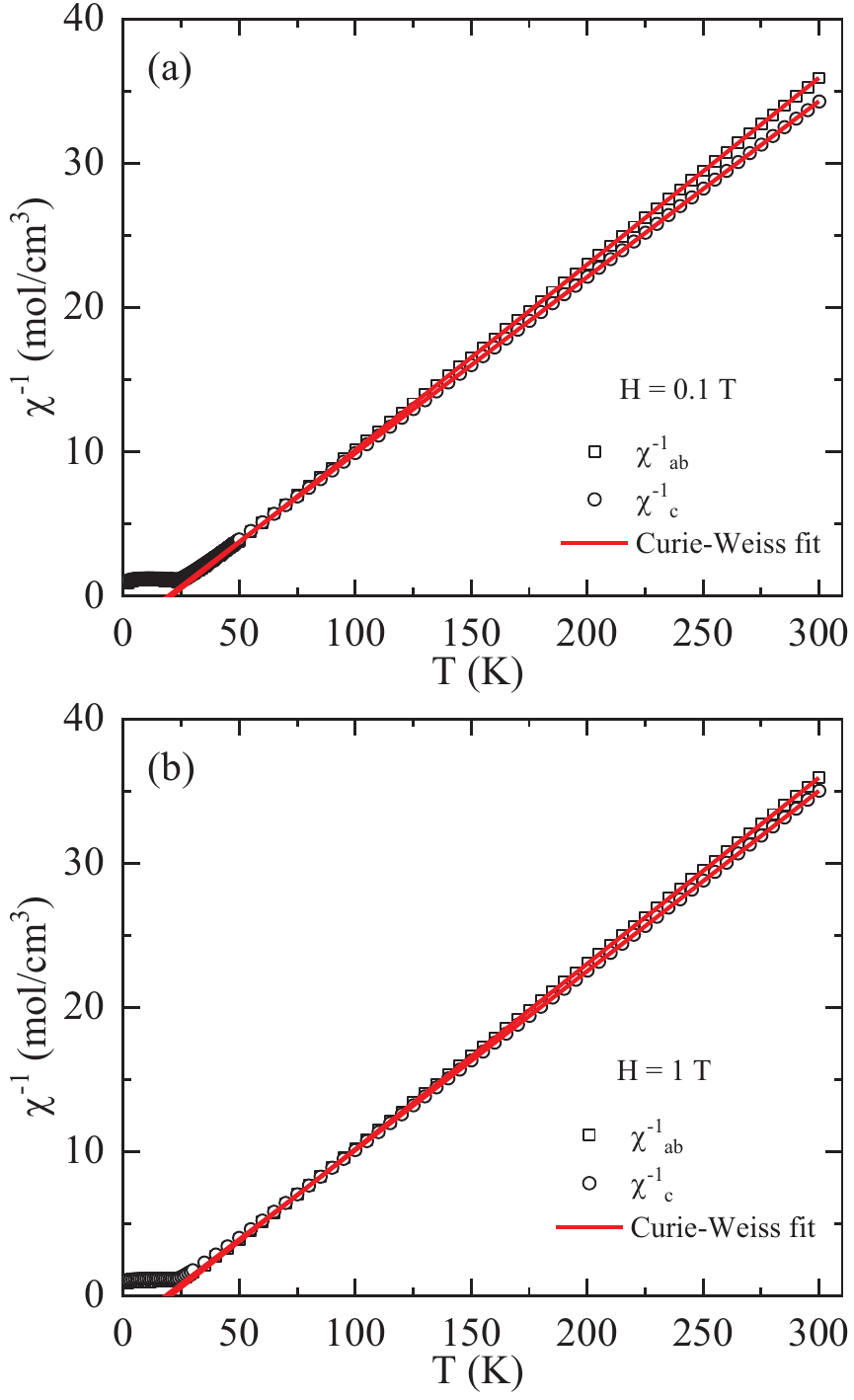}
\caption{Inverse magnetic susceptibility $\chi^{-1}(T)$ of \esa\ for (a)~$H = 0.1$~T and (b)~$H=1$~T, each with $H \parallel ab$ and $H \parallel c$, along with the respective  fits by the Curie-Weiss law~(\ref{Eq.CurieWeiss}).}
\label{Fig_CW_fit}
\end{figure}

The inverse susceptibilities $1/\chi_{ab}(T)$ and $1/\chi_c(T)$ of Crystal C measured in $H=0.1$~T are plotted in Fig.~\ref{Fig_CW_fit} for applied fields of (a)~0.1 and (b)~1~T\@. The data were analyzed in the paramagnetic state above $T_{\rm N}$ using the Curie-Weiss (CW) law
\bea
\chi_{\alpha}(T) = \frac{C_{\alpha}}{T-\theta_{\rm p\alpha}} \quad (\alpha =ab,\ c),
\label{Eq.CurieWeiss}
\eea
where $\theta_{\rm p\alpha}$ is the Weiss temperature with $H$ in the $\alpha$ direction.  The Curie constant $C_\alpha$  is given by
\bea
C_{\alpha}&=&\frac{N_{\rm A} {g_\alpha}^2S(S+1)\mu^2_{\rm B}}{3k_{\rm B}} = \frac{N_{\rm A}\mu^2_{\rm {eff, \alpha}}}{3k_{\rm B}},\\
\label{Eq.Cvalue1}
\mu_{\rm eff,\alpha}^2 &=& g_\alpha^2S(S+1)\mu^2_{\rm B}. \nonumber
\eea
where $N_{\rm A}$ is Avogadro's number, $g_\alpha$ is the spectroscopic-splitting factor ($g$ factor), $S=7/2$ is the Eu$^{2+}$ spin quantum number, $k_{\rm B}$ is Boltzmann's constant,  and $\mu_{\rm eff\alpha}$ is the effective moment of an Eu  spin.  It is evident from Fig.~\ref{Fig_CW_fit} that $\chi^{-1}(T)$ for $H = 0.1$ and 1~T are both well described by the CW law~\ref{Eq.CurieWeiss}) in the $T$ range 60--300~K for both $H \parallel ab$ and $H \parallel c$. The fitted parameters are summarized in Table~\ref{Tab.chidata}. The $\mu_{\rm {eff}\alpha}$ values obtained from the $C_\alpha$ values, $\mu_{\rm {eff\alpha}}$~($\mu_{\rm B}/{\rm Eu}) = \sqrt{8C_\alpha}$, are close to the value  7.94~$\mu_{\rm B}$/Eu expected for Eu$^{2+}$ spins with $S$ = 7/2 and $g = 2$. The presence of dominant FM interactions between the Eu spins is reflected in the positive values of $\theta_{\rm {p}\alpha}$.  The FM interactions are responsible for the FM alignment of the Eu spins in the $ab$~plane of the A-type AFM structure.  The values of $f_\alpha = \theta_{\rm p\alpha}/T_{\rm N}$ are listed for later use.

\begin{table}
\caption{\label{Tab.chidata} Listed are the applied magnetic field $H$, field direction, molar Curie constant $C_\alpha$ \mbox{($\alpha = ab, c$)}, effective moment per Eu spin $\mu{\rm_{eff\alpha}(\mu_B/Eu)} = \sqrt{8C_\alpha}$ and Weiss temperature $\theta\rm_{p\alpha}$ obtained from the $\chi^{-1}(T)$ data for \mbox{$H = 0.1$} and 1~T in Fig.~\ref{Fig_CW_fit} from fits by Eq.~(\ref{Eq.CurieWeiss}) in the temperature range  60--300~K\@.  Also listed are the ratios $f_\alpha = \theta_{\rm p\alpha}/T_{\rm N}$, where $T_{\rm N} = 24$~K\@.}
\begin{ruledtabular}
\begin{tabular}{cccccc}	
$H$ &Field &   $C_{\alpha}$ 		    &  $\mu_{\rm eff\alpha}$ 	& $\theta_{\rm p\alpha}$  & $f_\alpha$\\
(T) & direction	&	& $\rm{\left(\frac{cm^3 K}{mol}\right)}$    & $\rm{\left(\frac{\mu_B}{mol}\right)}$& (K)  \\
\hline
0.1	& $H \parallel ab$ 		& 7.84(1)	&	7.91(1)		& 20.8(1)		& 0.87	\\
0.1	& $H \parallel c$ 		& 7.91(1)	&	7.95(1)		& 19.2(2)		& 0.80 	\\
1	& $H \parallel ab$ 		& 7.85(1)	&	7.92(1)		& 20.5(1)		& 0.85	\\
1	& $H \parallel c$ 		& 7.90(1)	&	7.97(1)		& 18.3(1)		& 0.76 	\\
\hline
\end{tabular}
\end{ruledtabular}
\end{table}

\subsection{\label{Sec:Magsuscep} Low-temperature magnetic susceptibility}

\begin{figure}
\includegraphics[width = 3.3in]{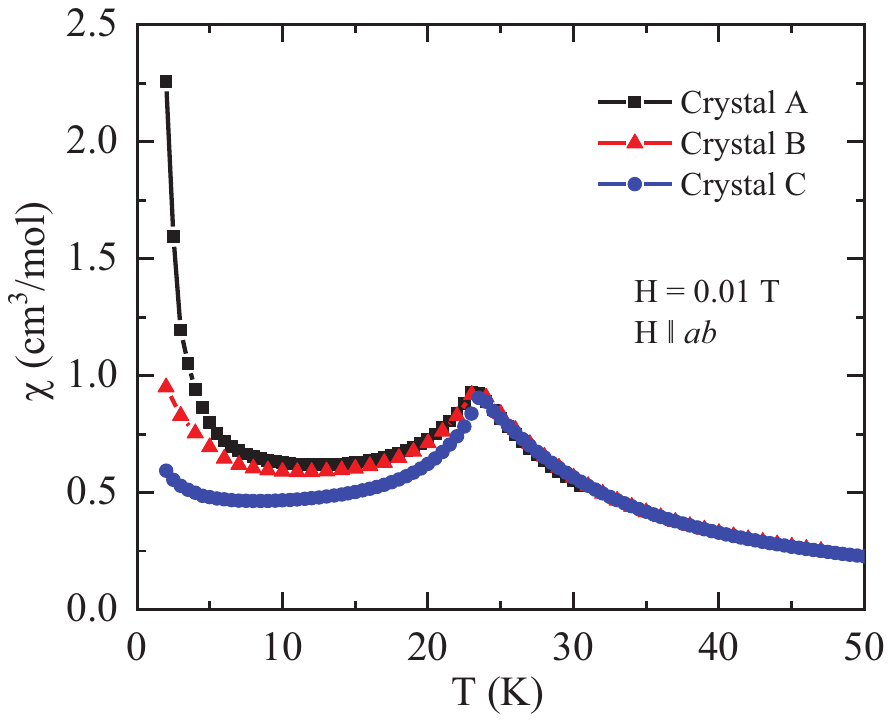}
\caption{Temperature $T$ dependence of the in-plane ($H \parallel ab$) magnetic susceptibility $\chi \equiv M/H$ measured in $H = 0.01$~T for three crystals (Crystal A, Crystal B, and Crystal C) grown in three different ways.  As discussed in Sec.~\ref{Sec:ExpDet}, the concentration of magnetic defects is expected to decrease from Crystal A to Crystal B to Crystal C, as observed. }
\label{Fig_Compare_M-T_diff_growth_100Oe}
\end{figure}

The temperature-dependent $\chi_{ab}(T) = M_{ab}(T)/H$ ($H \parallel ab$)  measured at $H = 0.01$~T for the three differently-grown crystals A, B, and C are shown in Fig.~\ref{Fig_Compare_M-T_diff_growth_100Oe}. Our data are consistent with those reported earlier~\cite{Li2019_EuSn2As2, Arguilla2017_EuSn2As2, Chen2020_EuSn2As2}, where  \esa\ crystals exhibit an AFM transition at $T_{\rm N} = 24$~K revealed by a sharp peak in $\chi_{ab}(T)$.  In addition, we find low-$T$ Curie-like upturns in our data below 10 K, which is strongest  for Crystal~A, moderate for Crystal~B and small for Crystal~C\@. It was previously suggested that the upturn in \esa\ crystals arises from an in-plane ferromagnetic component associated with canting of the magnetic moments~\cite{Li2019_EuSn2As2}. However, in this case, $\chi_{ab}(T)$ should not be sample-dependent as found here.  In order to obtain the intrinsic $\chi(T)$ at low~$T$, the Curie-like contribution must be corrected for as discussed below.

\begin{figure}
\includegraphics[width = 3.in]{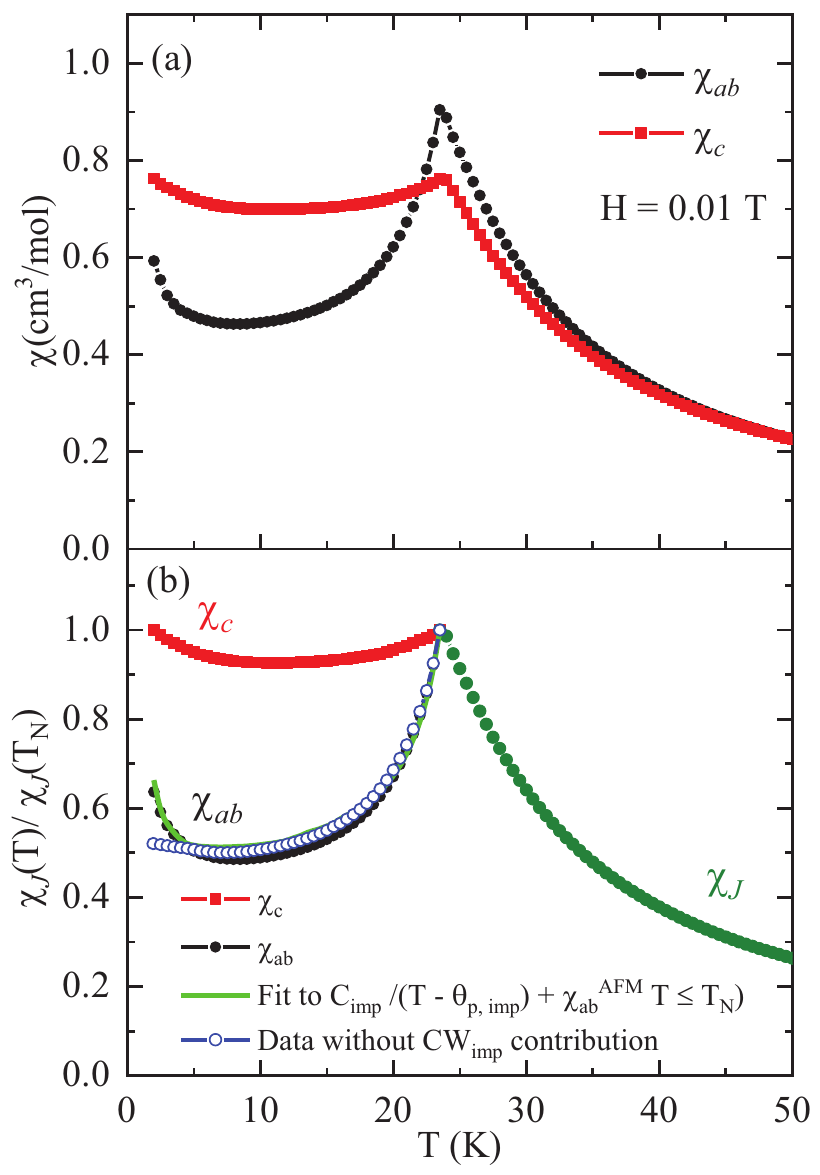}
\caption{(a) Temperature dependence of the zero-field-cooled in-plane ($\chi_{ab}, H \parallel ab$) and out-of-plane ($\chi_{c}, H \parallel c$) magnetic susceptibilities $\chi_\alpha = M_\alpha/H$ of \esa\ (Crystal C) measured in a magnetic field $H = 0.01$~T\@. The values of the anisotropic susceptibilities at $T_{\rm N} = 24$~K are \mbox{$\chi_{ab}(T_{\rm N}) = 0.904~{\rm cm^3/mol}$} and $\chi_c(T_{\rm N}) = 0.762~{\rm cm^3/mol}$. (b)~The Heisenberg susceptibility $\chi_J(T)$ in the paramagnetic state ($T > T_{\rm N}$) calculated by taking the spherical average of the data in panel (a) at $T > T_{\rm N}$ as shown by the green symbols on a scale normalized to that at $T = T_{\rm N}$\@.  $\chi_{ab}(T)$ (filled black circles) and $\chi_{c}(T)$ (filled red squares) with $T \leq T_{\rm N}$  are obtained by vertical translations of the respective $\chi$ data in panel~(a) to match the $\chi_J(T=T_{\rm N})$ data. The solid green curve gives the fit by Eq.~(\ref{Eq.Chi_ab_total}) to the $\chi_{ab}(T \leq T_{\rm N})$ data to obtain the low-$T$ magnetic-impurity Curie-Weiss contribution.  The open circles give $\chi_{ab}/\chi(T_{\rm N})$ vs $T$ after subtraction of the Curie-Weiss contribution which is consistent with A-type AFM ordering with the ordered moments ferromagnetically aligned in the $ab$~plane with the moments in adjacent planes along the $c$~axis aligned antiferromagnetically.}
\label{Fig_M-T_100Oe}
\end{figure}

The $\chi(T)$ was measured for an \esa\ crystal~C under both zero-field-cooled (ZFC) and field-cooled (FC) protocols and no thermal hysteresis was observed down to the lowest measured temperature of 1.8~K\@. Figure~\ref{Fig_M-T_100Oe}(a) shows the $T$ dependence of the ZFC in-plane ($\chi_{ab}$) and out-of-plane ($\chi_c$) data for Crystal C measured in a small $H = 0.01$~T (100 Oe) field.  A sharp peak associated with the AFM transition is observed at $T_{\rm N} = 24$~K for both $\chi_{ab}$ and $\chi_c$, in agreement with the earlier data. The low-$T$ upturn in the $\chi_{ab}$ data is larger than in the $\chi_c$ data, suggesting the presence of XY anisotropy for the magnetic impurity.

The $\chi_{ab}(T)$ and $\chi_c(T)$ data in Fig.~\ref{Fig_M-T_100Oe}(a) exhibit significant anisotropy in the paramagnetic temperature range $T_{\rm N} < T \lesssim 40$~K that likely arises from differences in the spin-fluctuation amplitudes in these two directions. For further analysis, the data were spherically averaged over this $T$ range yielding
\bea
\chi_{J}(T\geq T_{\rm N})=\frac{1}{3}[2\chi_{ab}(T)+\chi_{c}(T)].
\label{Eq.Heisenbergsusceptibility}
\eea
These $\chi_J(T)$ data normalized to $\chi_J(T = T_{\rm N})$ are plotted in Fig.~\ref{Fig_M-T_100Oe}(b), where the data for $\chi_{ab}$ and $\chi_{c}$ for $T \leq T_{\rm N}$ are obtained by vertical translations of the respective $\chi$ data to agree with $\chi_J(T = T_{\rm N})$. The $\chi_c(T)$ below $T_{\rm N}$ is nearly independent of~$T$ as expected from molecular-field theory (MFT) for crystallographically-equivalent Heisenberg ordered moments aligned in the $ab$~plane~\cite{Johnston2012, Johnston2015}.  On the other hand, the $\chi_{ab}(T)$ data decrease rapidly and attain a minimum value yielding the ratio \mbox{$\chi_{ab,\rm min}/\chi(T_{\rm N})\approx 1/2$}, which is indeed the value expected for an A-type Heisenberg antiferromagnet on a triangular lattice with moments aligned in the $ab$~plane within equally-populated three-fold domains~\cite{Pakhira2021_EuMg2Bi2}.

We utilize the MFT in Refs.~\cite{Johnston2012, Johnston2015} to calculate the low-field $\chi_{ab}(T\leq T_{\rm N})$.  Taking  \mbox{$f=\theta_{\rm p\,ave}/T_{\rm N}=0.83$ for $H=0.1$~T}  from Table~\ref{Tab.chidata}, we have
\bse
\label{Eqs:Chixy}
\be
\frac{\chi_{Jab}(T \leq T_{\rm N})}{\chi_J(T_{\rm N})}=  \frac{(1+\tau^*+2f+4B^*)(1-f)/2}{(\tau^*+B^*)(1+B^*)-(f+B^*)^2},
\label{eq:Chi_plane}
\ee
where
\be
B^*=  2(1-f)\cos(kd)\,[1+\cos(kd)] - f,
\label{eq:Bstar}
\ee
\be
t =\frac{T}{T_{\rm N}},\quad \tau^*(t) = \frac{(S+1)t}{3B'_S(y_0)}, \quad y_0 = \frac{3\bar{\mu}_0}{(S+1)t},
\ee
the ordered moment versus $T$ for $H=0$ is denoted by $\mu_0$, the reduced ordered moment $\bar{\mu}_0 = \mu_0/\mu_{\rm sat}$ is determined by numerically solving the self-consistency equation
\be
\bar{\mu}_0 = B_S(y_0),
\label{Eq:barmuSoln}
\ee
$B'_S(y_0) = [dB_S(y)/dy]|_{y=y_0}$, and the Brillouin function $B_S(y)$ is
\be
B_S(y)= \frac{1}{2 S}\left\{(2S+1){\rm coth}\left[(2S+1)\frac{y}{2}\right]-{\rm coth}\left(\frac{y}{2}\right)\right\}.
\ee
\ese
For $T=0$, using Eq.~(\ref{eq:Chi_plane}) yields~\cite{Johnston2012, Johnston2015}
\be
\label{Eq:kd}
\frac{\chi_{Jab}(T=0)}{\chi_{Jab}(T_{\rm N})}=\frac{1}{2[1+2~{\rm cos}(kd)+2~{\rm cos}^2(kd)]}
\ee
Thus for an A-type antiferromagnet with $kd\to180^\circ$ and the moments aligned in the $ab$~plane, one obtains a value of~1/2 for this ratio, in agreement with the above value \mbox{$\chi_{ab,\rm min}/\chi(T_{\rm N})\approx 1/2$}.

Now we fit  $\chi_{ab}(T\leq T_{\rm N})$ in  Fig.~\ref{Fig_M-T_100Oe}(b) by the sum of the AFM contribution described by Eqs.~(\ref{Eqs:Chixy}) and a Curie-Weiss magnetic-defect contribution according to
\bea
\chi_{ab}(T \leq T_{\rm N}) = \chi_{Jab}(T \leq T_{\rm N}) + \frac{C_{\rm imp}}{T-\theta_{\rm p\,imp}},
\label{Eq.Chi_ab_total}
\eea
where $C_{\rm imp}$ and $\theta_{\rm p\,imp}$ are the Curie constant and Weiss temperature associated with the magnetic impurities/defects discussed previously.  The best fit of the data by Eq.~(\ref{Eq.Chi_ab_total}) is shown in Fig.~\ref{Fig_M-T_100Oe}(b) as the green solid curve with $C_{\rm imp} = 0.0874$~cm$^3$-K/mol~f.u.\ (f.u.\ means formula unit) and $\theta_{\rm p\,imp} = 1.4$~K\@.  The blue open circles show the data after subtracting the low-$T$ CW contribution.  The $C_{\rm imp}$ value is only $\approx 1.1$\% of the value $\rm{7.88\,cm^3}$\,K/mol for Eu$^{2+}$ with $g=2$ and $S = 7/2$. From Fig.~\ref{Fig_M-T_100Oe}(b), after correcting for the CW contribution the data exhibit \mbox{$\chi_{ab}(T = 1.8~{\rm K})/\chi_{ab}(T = T_{\rm N}) \approx 1/2$,} as expected for A-type AFM order with turn angle $kd \to 180^\circ$.

\begin{figure}
\includegraphics[width = 3.in]{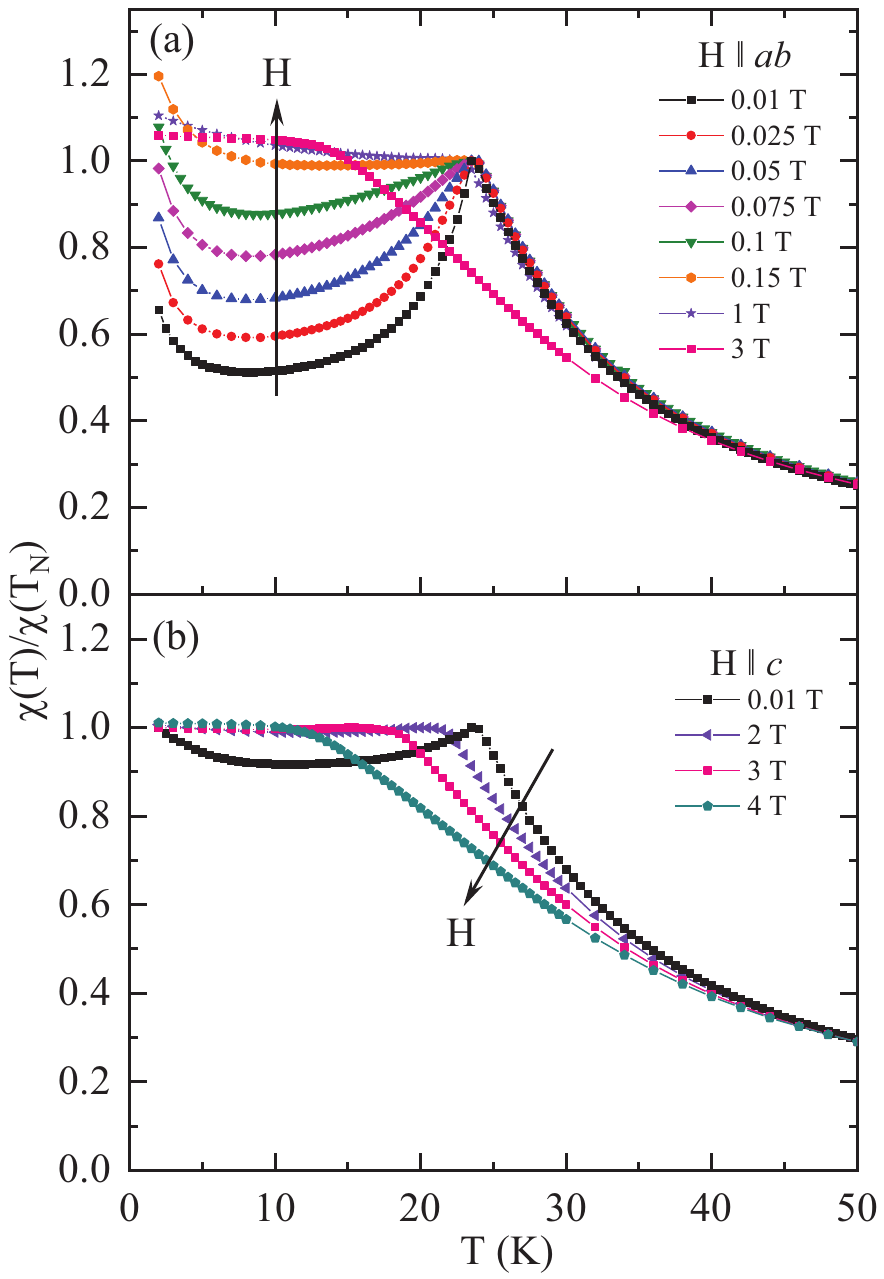}
\caption{Temperature dependence of the magnetic susceptibilities (a)~$\chi_{ab}(H)\equiv M_{ab}(H)/H$ and (b)~$\chi_c(H)\equiv M_c(H)/H$ for Crystal C normalized by the respective values at $T = T_{\rm N}(H)$ in applied magnetic fields $H$ from 0.01~T to 0.15~T\@.  The $\chi_{ab}(H)$ data are strongly dependent on $H$, while the $\chi_c(H)$ data are almost independent of $H$.  $T_{\rm N}$ is seen to decrease by about 40\% at the highest field.  The data suggest a smooth evolution of the magnetic structure on increasing $H$ from 0.01 to $\sim 0.15$T\@.  }
\label{Fig_M-T_diff_fields}
\end{figure}

The $\chi_{ab}(T)$ and $\chi_{c}(T)$ data normalized by the respective values at $T_{\rm N}$ for different applied magnetic fields are shown in Figs.~\ref{Fig_M-T_diff_fields}(a) and \ref{Fig_M-T_diff_fields}(b), respectively. Neglecting the low-$T$ upturns, the extrapolated \mbox{$\chi_{ab}(T = 1.8~{\rm K})/\chi_{ab}(T = T_{\rm N})$} value shows a smooth increase from $\approx 0.5$ as discussed above to larger values with increasing $H_{ab}$ and saturates at $\approx 1$, while the ratio  $\chi_{c}(T = 1.8~{\rm K})/\chi_{c}(T = T_{\rm N})$ is nearly independent of~$T$ and~$H$ for $T \leq T_{\rm N}$\@.

These $\chi_\alpha(H,T\leq T_{\rm N})$ results are similar to those we observed previously for the trigonal compound ${\rm EuMg_2Bi_2}$ containing stacked triangular lattices of Eu$^{2+}$ spins $S=7/2$~\cite{Pakhira2020} where the magnetic structure in $H=0$ was also found  by neutron-diffraction measurements to be A-type AFM  with the Eu moments aligned in the $ab$~plane~\cite{Pakhira2021}.  In particular, we find here that with increasing $ab$-plane field $M_{ab}$ becomes nearly independent of $H$ for $H\gtrsim 0.125$~T\@.  A similar result was interpreted in Ref.~\cite{Pakhira2021} as possibly arising from the spins rotating with increasing $H_{ab}$ to be perpendicular to $H$, which gives $\chi_{ab}(T\leq T_{\rm N})/\chi(T_{\rm N}) \approx \chi_{c}(T\leq T_{\rm N})/\chi(T_{\rm N}) \approx 1$ as also observed here.

The $ab$-plane anisotropy arising from magnetic-dipole interactions between moments on a triangular lattice is small and with a  strong preference of these interactions for $ab$-plane moment alignment over $c$-axis alignment~\cite{Johnston2016}.  For an isolated layer of Eu spins with the moments aligned ferromagnetically in the $ab$~plane, $ab$-plane alignment is more stable than $c$-axis alignment by $\Delta E = 0.294$~meV, which corresponds to a magnetic field anisotropy  $H_{\rm anis} = \Delta E/(7~\mu_{\rm B}) = 7.25$~kOe.  Thus the small $ab$-plane magnetic field of 0.15~T discussed above is insufficient to cause a spin-flop transition with the moments having a $c$-axis component at $T\ll T_{\rm N}$.

Therefore we propose that with increasing $ab$-plane field $H_{ab}$ at $T=2$~K, the above three AFM domains initially at 120$^\circ$ to each other as dictated by the Eu trigonal point symmetry $\bar{3}m$ begin to align perpendicular to $H_{ab}$, which according to Fig.~\ref{Fig_M-T_diff_fields}(a) is completed at $H_{ab}= 0.15$~T\@.  Then with a further increase in $H_{ab}$ the collinear spins in each domain progressively cant towards $H_{ab}$ until they reach FM alignment at the critical field $H_{ab}^{\rm c} = 3.90$~T listed in Table~\ref{Tab.Criticalfield}~\cite{Johnston2017}.

\subsection{Magnetization versus applied magnetic field isotherms}

\begin{figure}
\includegraphics[width = 3.in]{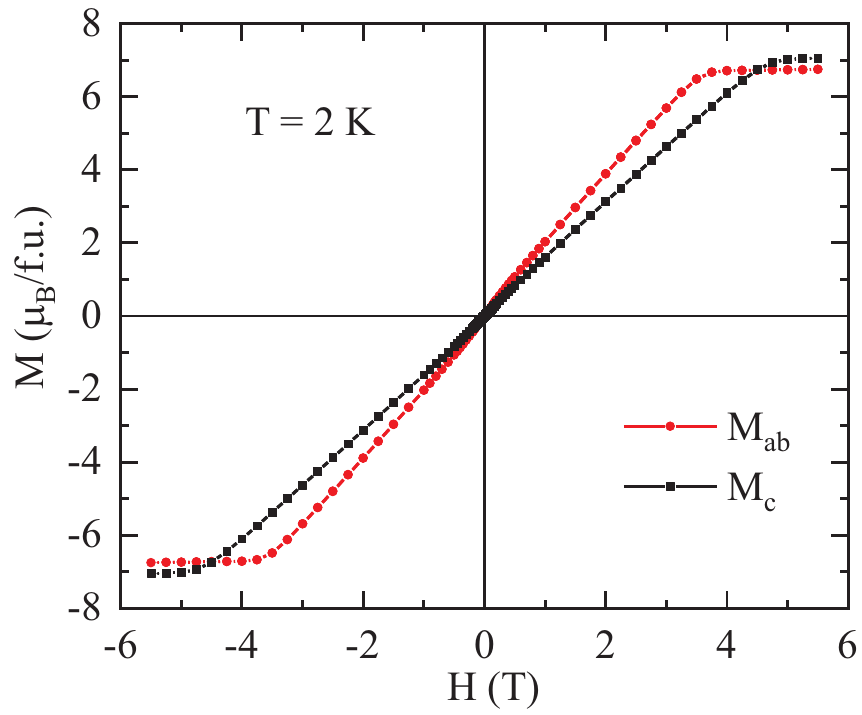}
\caption{Four-quadrant magnetization versus field $M(H)$ hysteresis data for Crystal C with $ab$-plane and $c$-axis magnetic-field orientations at $T = 2$~K\@. No hysteresis is observed for either field direction.  }
\label{Fig_M-H_hysteresis}
\end{figure}

\begin{figure*}
\includegraphics[width = 7in]{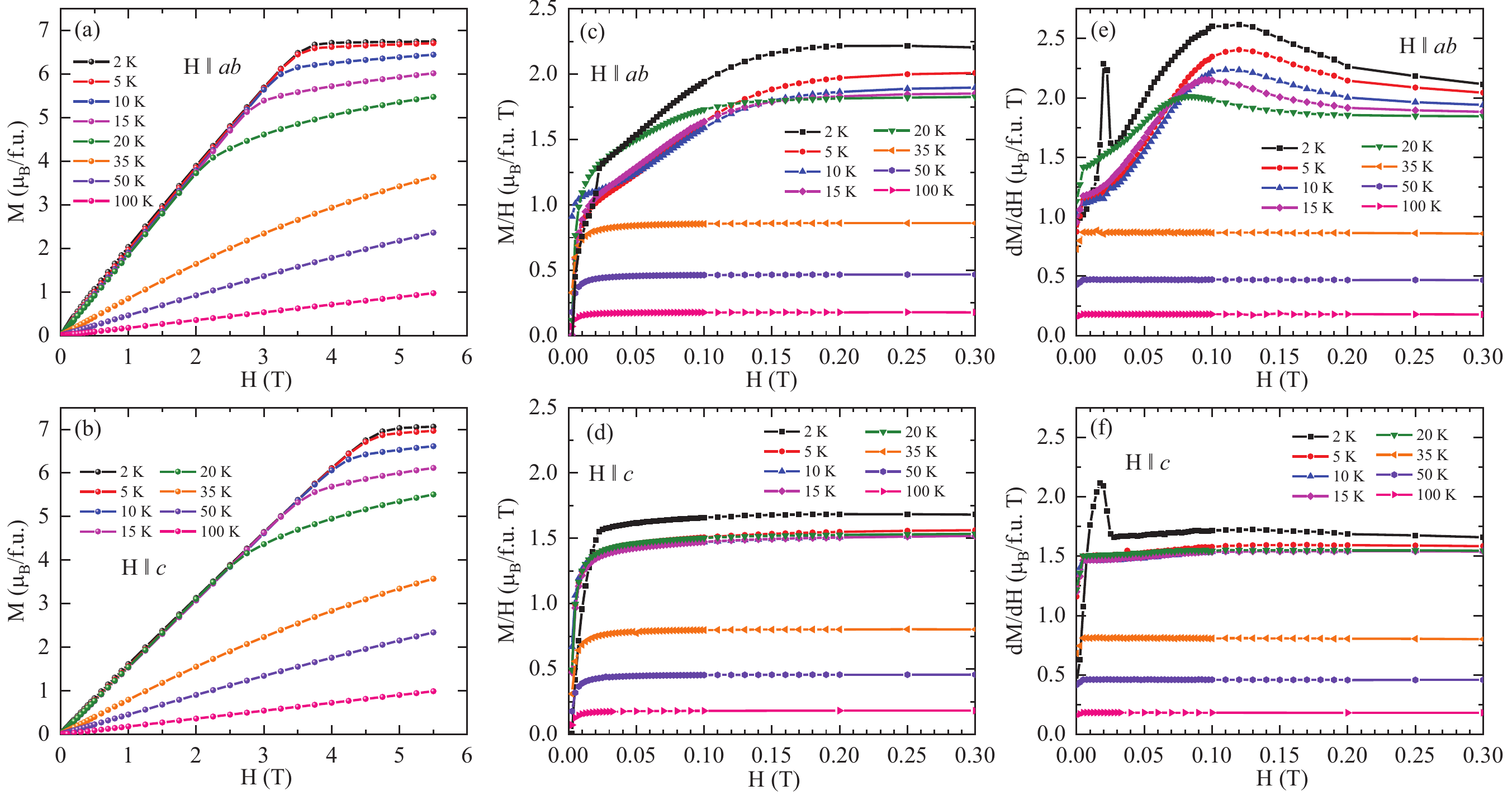}
\caption{(a, b) Magnetization $M$ isotherms for Crystal C as a function of applied magnetic field $H$. Low-field $H$-dependences of (c, d)~$M/H$, and (e, f) $(dM/dH)$ at different temperatures for $H \parallel ab$ and $H \parallel c$, respectively.}
\label{Fig_M-H_diff_temp}
\end{figure*}

The field dependence of the isothermal magnetization~$M$ was measured at different temperatures for both in-plane and out-of-plane field directions. Figure~\ref{Fig_M-H_hysteresis} shows four-quadant $M(H)$ hysteresis curves recorded at $T = 2$~K for $H \parallel ab$ and $H \parallel c$.  No evidence of magnetic hysteresis was observed. On the scale of the figure, the $M_{ab}(H)$ and $M_{c}(H)$ data are linear in field and then saturate to values of $\approx 6.8~\mu_{\rm B}$  and $\approx 7~\mu_{\rm B}$ at the critical fields $H_{ab}^{\rm c} = 3.90(5)$~T and $H_{c}^{\rm c} = 5.00(5)$~T, respectively. The $T$ dependence of $H_{ab}^{\rm c}$ and $H_{ab}^{\rm c}$ obtained from discontinuities in slope of $M_\alpha(H)$ isotherms in Figs.~\ref{Fig_M-H_diff_temp}(a) and \ref{Fig_M-H_diff_temp}(b), respectively, are listed in Table~\ref{Tab.Criticalfield}.

\begin{table}
\caption{\label{Tab.Criticalfield} The in-plane and out-of-plane critical fields $H_{ab}^{\rm c}$ and $H_{c}^{\rm c}$, respectively, at different temperatures obtained from Figs.~\ref{Fig_M-H_diff_temp}(a) and \ref{Fig_M-H_diff_temp}(b), respectively. The critical field for a given field direction is the field for which the slope of $M(H)$ shows a discontinuity with increasing field.}
\begin{ruledtabular}
\begin{tabular}{ccc}	
$T$~(K)			& $H_{ab}^{\rm c}$ (T)	 &  $H_{c}^{\rm c}$ (T) \\
 2	& 3.63(7)	 &  4.70(7) \\
5	& 3.56(7)	 &  4.60(7) \\
10	& 3.28(7)	 &  4.32(7) \\
15	& 3.0(1)	 &  3.7(1) \\
20	& 2.5(1)	 &  3.0(1) \\
\end{tabular}
\end{ruledtabular}
\end{table}

The $M_{ab}(H)$ and $M_{c}(H)$ data measured at different temperatures are plotted for $H=0$--5.5~T in Figs.~\ref{Fig_M-H_diff_temp}(a) and \ref{Fig_M-H_diff_temp}(b), respectively, and $M/H$ vs $H$ for $H \parallel ab$ and $H \parallel c$ in the low-field region $H=0$--0.3~T in Figs.~\ref{Fig_M-H_diff_temp}(c) and \ref{Fig_M-H_diff_temp}(d), respectively. The $M_{c}(H)/H$ vs $H$ data are almost independent of~$H$ except for $H \rightarrow 0$, which is attributed to the remanent field in our magnetometer.

The $M_{ab}(H)$ data in Figs.~\ref{Fig_M-H_diff_temp}(c,~e) exhibit a sharp transition-like feature at $H = 0.02$~T at the lowest temperature $T = 2$~K, whereas for the $M_{c}(H)$ data this signature is weak as revealed in Figs.~\ref{Fig_M-H_diff_temp}(d,~f). No signature of this anomaly is observed for $T \geq 5$~K\@. This feature is likely due to the field-induced saturation of the magnetic defects. These saturations give rise to a significant enhancement in the lowest-$T$ $\chi(T)$ causing a rapid upturn in the low-field region. In this case the value $H \approx 0.02$~T is the critical field value beyond which the magnetic defects tend to saturate. Since the defect contribution is larger in $M_{ab}$ compared to $M_{c}$, the jump in $dM_{ab}(H)/dH$ observed at $T = 2$~K is sharper for the $ab$-plane data than for the $c$-axis data.

The $M_{ab}(H)/H$ vs $H$ data at $T=2\,{\rm K}$ in Fig.~\ref{Fig_M-H_diff_temp}(c) deviate from a constant value for $H \lesssim 0.2$~T\@. This feature is reflected more clearly in the derivative plot $dM_{ab}(H)/dH$ vs $H$ in Fig.~\ref{Fig_M-H_diff_temp}(e).  A clear maximum is observed in $dM_{ab}(H)/dH$ at $H_{\rm p} \approx 0.12$~T for $T = 2$~K, where the maximum shifts to lower $H$  with increasing $T$, whereas no such feature is observed in the $dM_{c}(H)/dH$ data in Fig.~\ref{Fig_M-H_diff_temp}(f) in that $H$ region. This feature in the $dM_{ab}(H)/dH$ data only persists up to $T \approx T_{\rm N}$, indicating that  it likely arises from a field-induced change in the A-type ground-state spin configuration in $H=0$ as discussed at the end of the previous section.  Note that $H_{\rm p}(T=2\,{\rm K})$ is much smaller that the critical field $H_{ab}^{\rm c}(T=2\,{\rm K}) = 3.90$~T  in Table~\ref{Tab.Criticalfield}.

\section{\label{Cp} Heat capacity measurements}

\begin{figure}
\includegraphics[width =3.in]{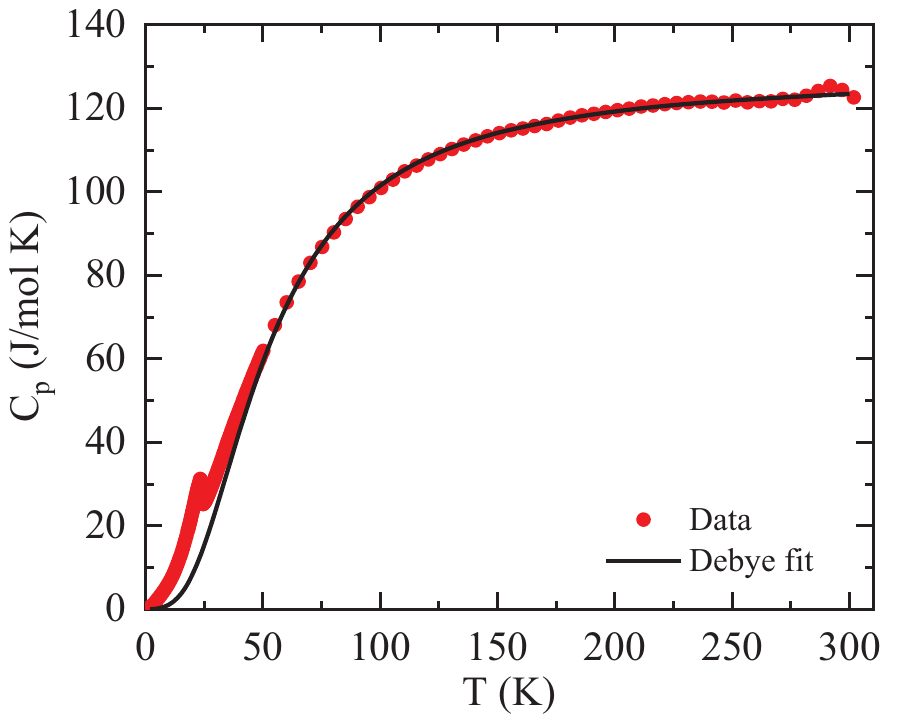}
\caption{Temperature dependence of zero-field heat capacity $C_{\rm p}$ for Crystal C along with a fit  by Eqs.~(\ref{Eqs:Debye_Fit}) in the temperature range 50--300~K\@. The data are fitted well for $T \geq 60$~K\@. The bump in the data at $\approx 290$~K is an experimental artifact.}
\label{Fig_Heat_capacity}
\end{figure}

\begin{figure*}
\includegraphics[width = 7in]{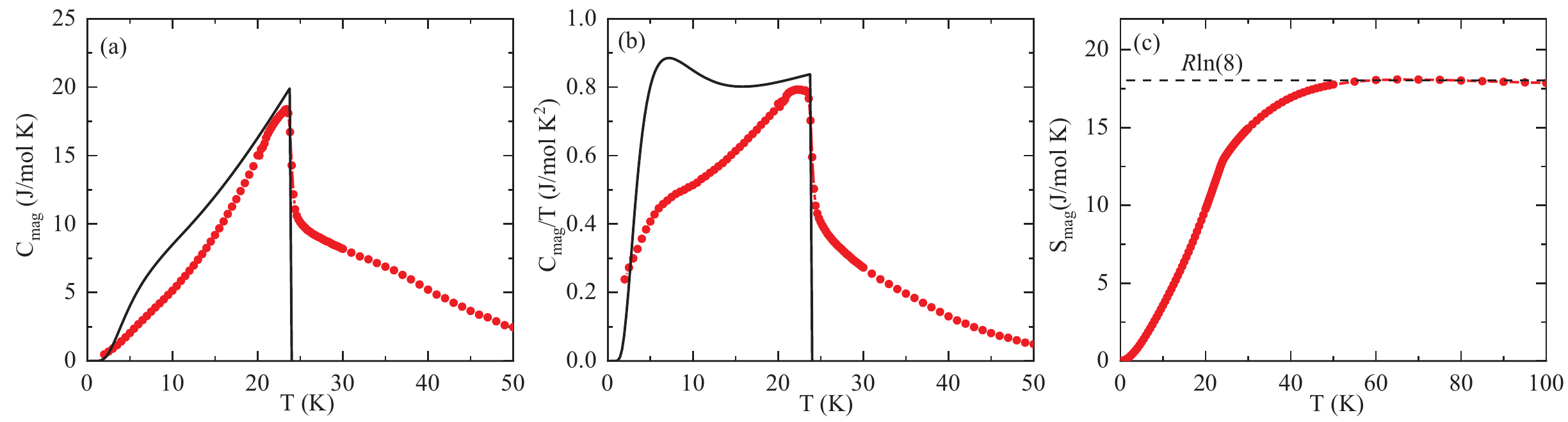}
\caption{Temperature $T$ dependence of (a) the magnetic component of the heat capacity $C_{\rm {mag}}$, (b) $C_{\rm {mag}}/T$ , and (c) the magnetic entropy $S_{\rm {mag}}$ obtained from the $C_{\rm {mag}}/T$ data in panel (b) using Eq.~(\ref{Eq:Entropy}) for \esa\ (Crystal C). The solid black lines in panels (a) and (b) represent the predictions of MFT in Eq.~(\ref{Eq:deltaCp_T_MFT}) for $S = 7/2$ and $T_{\rm N} = 24$~K\@. $S_{\rm {mag}}$ attains its saturation value $R\ln(2S+1) = R\ln8$ by $T \approx 60$~K, demonstrating the presence of AFM correlations above $T_{\rm N} = 24$~K.}
\label{Fig_Heat_capacity_MFT}
\end{figure*}

The temperature dependence of the zero-field heat capacity $C_{\rm p}(T)$ for an \esa\ crystal~C is shown in Fig.~\ref{Fig_Heat_capacity} in the temperature range 1.8--300~K\@. A $\lambda$-type peak is observed at $T_{\rm N} = 24$~K associated with the AFM ordering. The $C_{\rm p}(T)$ value of $\approx 123$~ J/mol~K at $T = 300$~K  is close to the classical high-$T$ Dulong-Petit limit $C_{\rm V} = 3nR = 124.71$ J/mol K associated with lattice vibrations for this compound, where $n=5$ is the number of atoms per formula unit and $R$ is the molar-gas constant. The bump in the $C_{\rm p}(T)$ around $T = 290$~K is an instrumental artifact arising from melting of the Apiezon N grease used for making thermal contact between the crystal and sample platform of the heat capacity puck.  We fitted the $C_{\rm p}(T)$ data in the measured temperature range using the sum of a Sommerfeld electronic heat capacity contribution $\gamma T$ and the Debye model for the lattice heat capacity given by
\bse
\label{Eqs:Debye_Fit}
\bea
C_{\rm p}(T) &=& \gamma T+ n~ C_{\rm V\,Debye}(T/\Theta_{\rm D}), \\*
C_{\rm V\,Debye}(T) &=& 9R \left(\frac{T}{\Theta_{\rm D}}\right)^3\int_{0}^{\Theta_{\rm D}/T}\frac{x^4e^x}{(e^x-1)^2} dx,
\label{Eq:CVDebye}
\eea
\ese
where $\gamma$ is Sommerfeld electronic heat capacity coefficient and $\Theta_{\rm D}$ is the Debye temperature. The lattice heat capacity $C_{\rm V\,Debye}(T)$  in Eq.~(\ref{Eq:CVDebye}) was parameterized using an accurate Pad\'{e} approximant function~\cite{Goetsch_2012}.  A fit of the data in Fig.~\ref{Fig_Heat_capacity} in the temperature range 60--300~K by Eqs.~(\ref{Eqs:Debye_Fit}) yielded \mbox{$\gamma$ = 5.9(6)~mJ/(mol~K$^2$)} and \mbox{$\Theta_{\rm D}$ = 210(1)~K}\@.

The magnetic contribution $C_{\rm {mag}}(T)$ to the heat capacity is estimated by subtracting the lattice contribution given by Eq.~(\ref{Eq:CVDebye}) with the fitted value of $\Theta_{\rm D}$ from the zero-field $C_{\rm p}(T)$ data. The $C_{\rm {mag}}(T)$ is plotted in Fig.~\ref{Fig_Heat_capacity_MFT}(a) which shows a clear $\lambda$-type peak at $T_{\rm N} = 24$~K followed by an extended tail in the higher-$T$ region, indicating the persistence of dynamic short-range AFM ordering at $T>T_{\rm N}$\@. This feature has also been observed in other  Eu$^{2+}$ $S=7/2$ AFM compounds, {\it e.g.}, \cite{Pakhira2020_EuMg2Bi2, Sangeetha_EuCo2As2_2018, Sangeetha_EuCo2P2_2016, Sangeetha_EuNi2As2_2019}. 

The origin of the strong AFM correlations in \esa\ above $T_{\rm N}$ may be traced to the very soft response of $M_{ab}(H)$ at low fields in Figs.~\ref{Fig_M-H_diff_temp}(c, e).  Due to the presence of the strong short-range AFM ordering above $T_{\rm N}$, the heat-capacity jump \mbox{$\Delta C_{\rm {mag}}(T_{\rm N})\approx 10$~J/mol~K} in Fig.~\ref{Fig_Heat_capacity_MFT}(a) is much smaller than $\Delta C_{\rm {mag}}(T_{\rm N})$ for $S=7/2$ expected from MFT given by~\cite{Johnston2015}
\bea
\label{Eq:deltaCp}
\Delta C_{\rm_{mag}}^{\rm MFT} = R\frac{5S(S+1)}{1+2S+2S^2} = 20.1~{\rm J/(mol~K)}.
\eea

The MFT prediction of $C_{\rm {mag}}(T)$ is given by~\cite{Johnston2015}
\bea
\label{Eq:deltaCp_T_MFT}
\frac{C_{\rm {mag}}(t)}{R} = \frac{3S\bar{\mu}_0^2(t)}{(S + 1)t[\frac{(S + 1)t}{3B^{\prime}_S(t)} - 1]},
\eea
where the parameters are defined in Eqs.~(\ref{Eqs:Chixy}). Figures~\ref{Fig_Heat_capacity_MFT}(a) and~\ref{Fig_Heat_capacity_MFT}(b) depict $C_{\rm {mag}}$ and $C_{\rm {mag}}/T$ vs $T$ along with the MFT predictions.  Dramatic differences between the MFT predictions and the experimental data are seen, again attributed to the strong influence of dynamic short-range AFM fluctuations.

The magnetic entropy $S_{\rm {mag}}(T)$ for \esa\ is obtained from the $C_{\rm {mag}}(T)/T$ data  in Fig.~\ref{Fig_Heat_capacity_MFT}(b) according to
\bea
\label{Eq:Entropy}
S_{\rm mag}(T) = \int_{0}^{T}\frac{C_{\rm {mag}}(T^\prime)}{T^\prime} dT^\prime,
\eea
where the $C_{\rm {mag}}(T)/T$ data in the $T$ range 0--1.8~K are extrapolated using the MFT prediction shown in Fig.~\ref{Fig_Heat_capacity_MFT}(b). As can be seen from Fig.~\ref{Fig_Heat_capacity_MFT}(c), the temperature dependence of zero-field $S_{\rm {mag}}$ saturates at a value of \mbox{$\approx 18$ J/mol K,} which is close to the theoretical high-$T$ limit $S_{\rm mag} = R{\rm ln}(2S + 1) = 17.29$~J/(mol~K) expected for \esa\ with $S = 7/2$. The $S_{\rm {mag}}(T)$ saturates around $T = 60$~K which is more than twice $T_{\rm N}=24$~K of the compound. This again reflects the strong dynamic short-range AFM ordering that occurs above $T_{\rm N}$.  Similar observations of entropy release above $T_{\rm N}$ were found for other Eu$^{2+}$ and Gd$^{3+}$ compounds with spins~7/2 in Refs.~\cite{Sangeetha_EuCo2As2_2018, Sangeetha_EuCo2P2_2016, Sangeetha_EuNi2As2_2019, Anand2015, Anand2014, Singh2009, Pandey2017, Johnston2021}.

\begin{figure}
\includegraphics[width = 3.in]{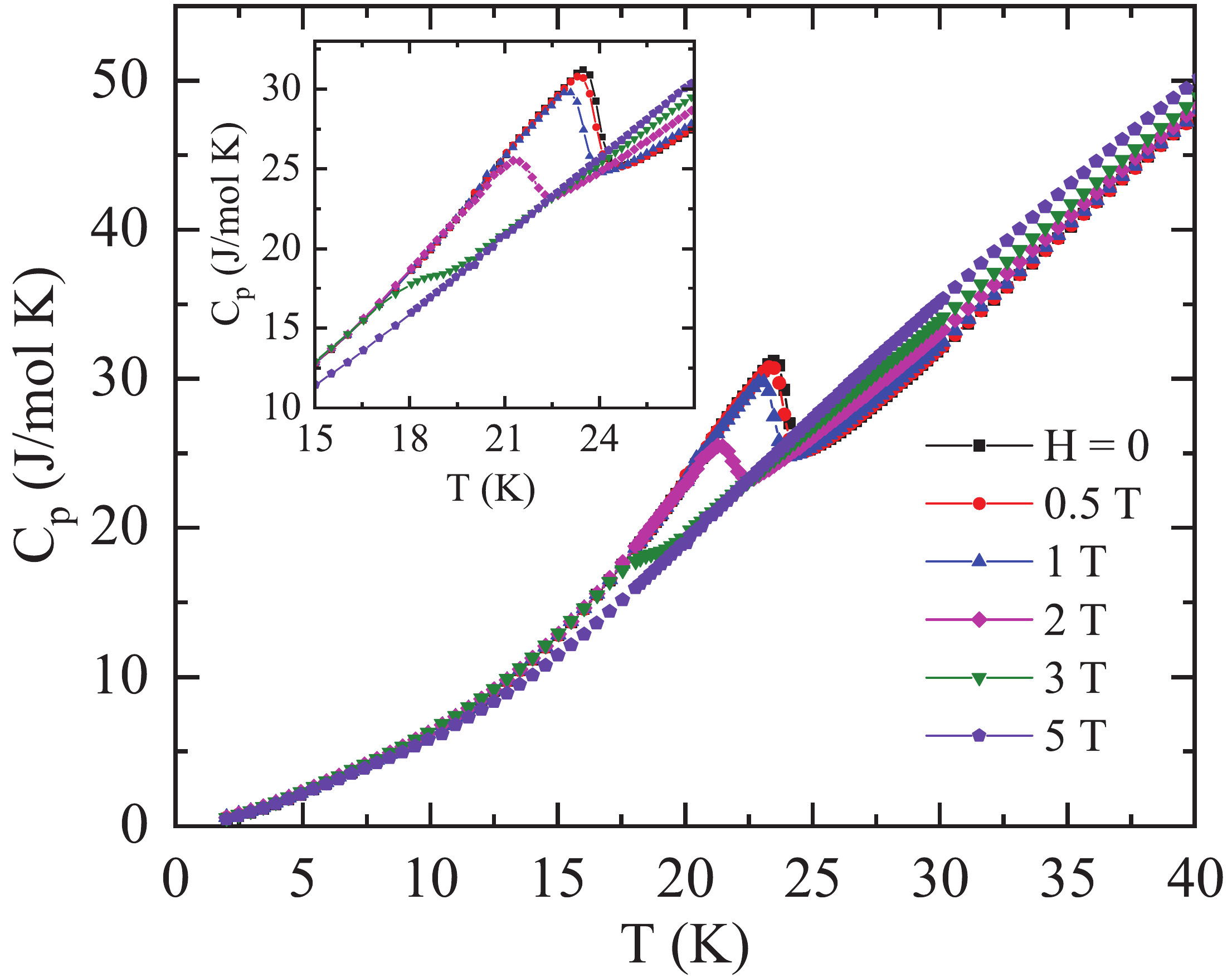}
\caption{$C_{\rm p}(T)$ of Crystal C measured for different applied magnetic fields with $H \parallel c$ showing the suppression of the AFM ordering temperature $T_{\rm N}$ with increasing $H$\@. Inset: expanded plot of the data between 15 and 27~K\@.}
\label{Fig_Heat_capacity_Field}
\end{figure}

The $C_{\rm p}(T)$ data measured at different applied magnetic fields with $H\parallel c$ are plotted in Fig.~\ref{Fig_Heat_capacity_Field}.  The data show that $T_{\rm N}$ shifts to lower $T$ with increasing~$H$ as expected. These data are used later in Sec.~\ref{Summary} to help construct the magnetic phase diagram in the $H$-$T$ plane.

\section{\label{rho} Electrical resistivity measurements}

\begin{figure}[h]
\centering
\includegraphics[width = 3.3in]{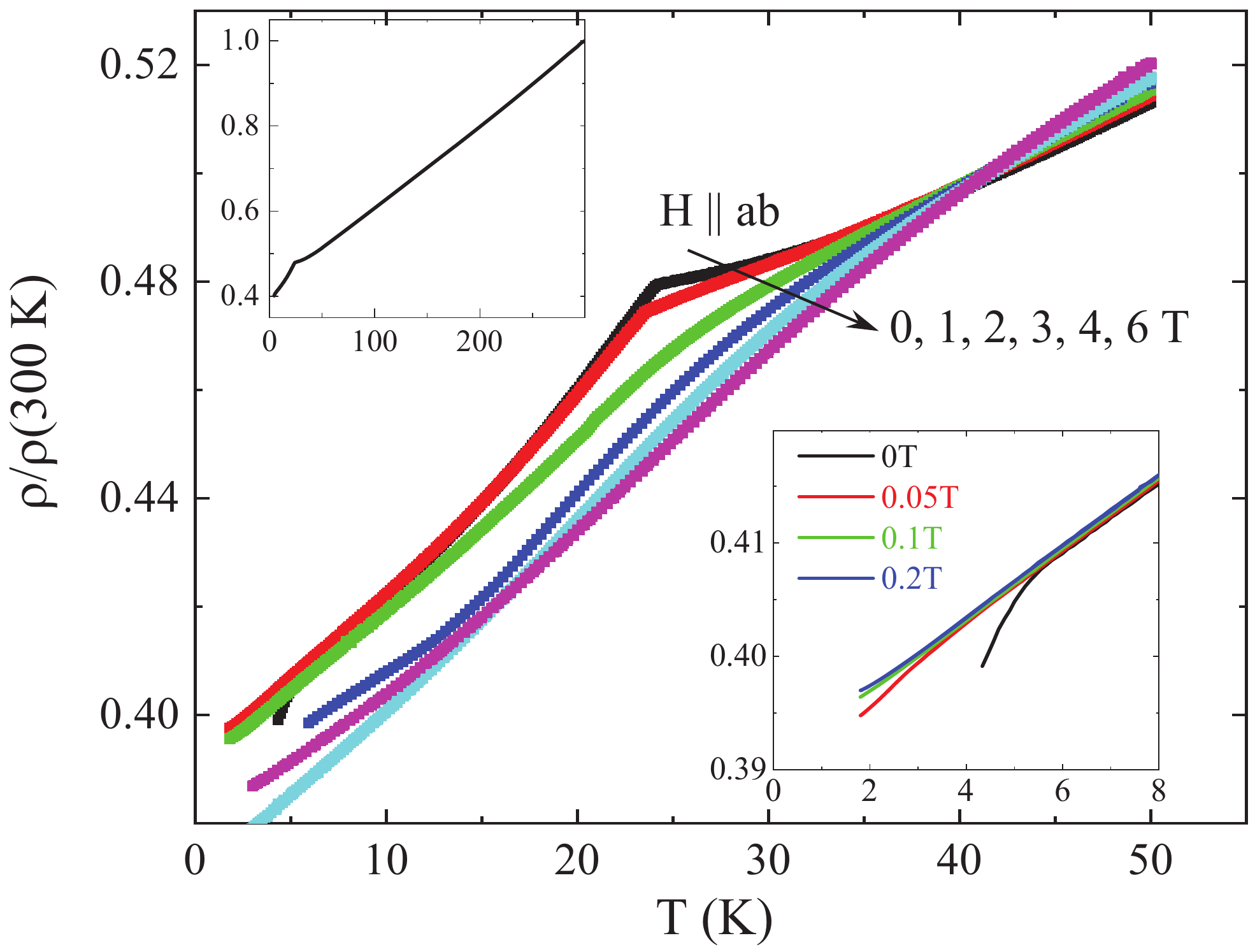}
    \caption{In-plane electrical resistivity of EuSn$_2$As$_2$, presented using a normalized $\rho/\rho$(300~\rm K) scale. The $T$ dependence over the whole $T$ range in $H=0$ is shown in the top-left inset. The main panel zooms in on the $T$ range of the main AFM transition, with measurements made in $H{\parallel ab}$ of 0~T (black, top curve), 1~T (red), 2~T  (green), 3~T (blue), 4~T (cyan) and 6~T (magenta). Note the negative magnetoresistance in the $T$ range below 40~K and the crossover to positive magnetoresistance above. The bottom-right inset shows a zoom of the second feature in the $T$-dependent resistivity, a slight decrease below $T_L \sim 6$~K in $H=0$ (black curve). This feature broadens in $H= 0.05$~T (red curve) and is suppressed with $H= 0.1$~T (green curve) and 0.2~T (blue curve).}
\label{resistivity}
\end{figure}

\begin{figure}
\centering
\includegraphics[width = 3.3in]{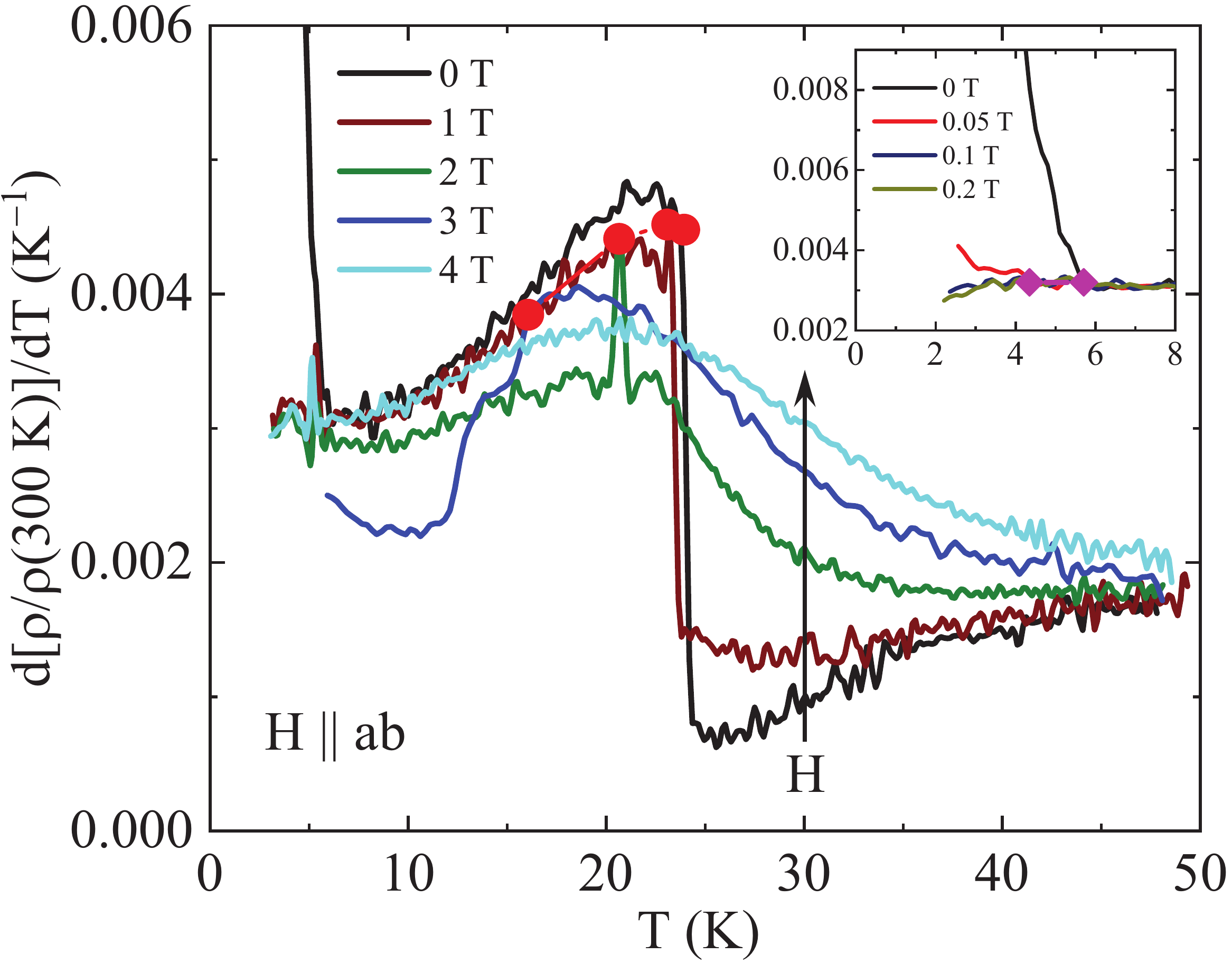}
\caption{Temperature $T$ derivative $d[\rho(T)/\rho(300~{\rm K})]/dT$ of EuSn$_2$As$_2$ up $T=50$~K\@. The derivative shows a jump at $T_{\rm N} \approx 23$~K in zero field, with the maximum denoted by a filled red circle. The jump remains sharp in a magnetic field $H \parallel ab$ of 1~T (maroon  curve), and an additional sharp peak occurs for 2~T and a step at 3~T with a broad crossover background, while the sharp peak increases in magnitude.  The feature at $T_{\rm N}$ further broadens in 4~T\@. The inset shows a suppression of the zero-field anomaly in the derivative at 6~K with magnetic field, and magenta diamonds show the position of the feature onset. }
\label{derivative}
\end{figure}

\begin{figure}
\centering
\includegraphics[width=3.4in]{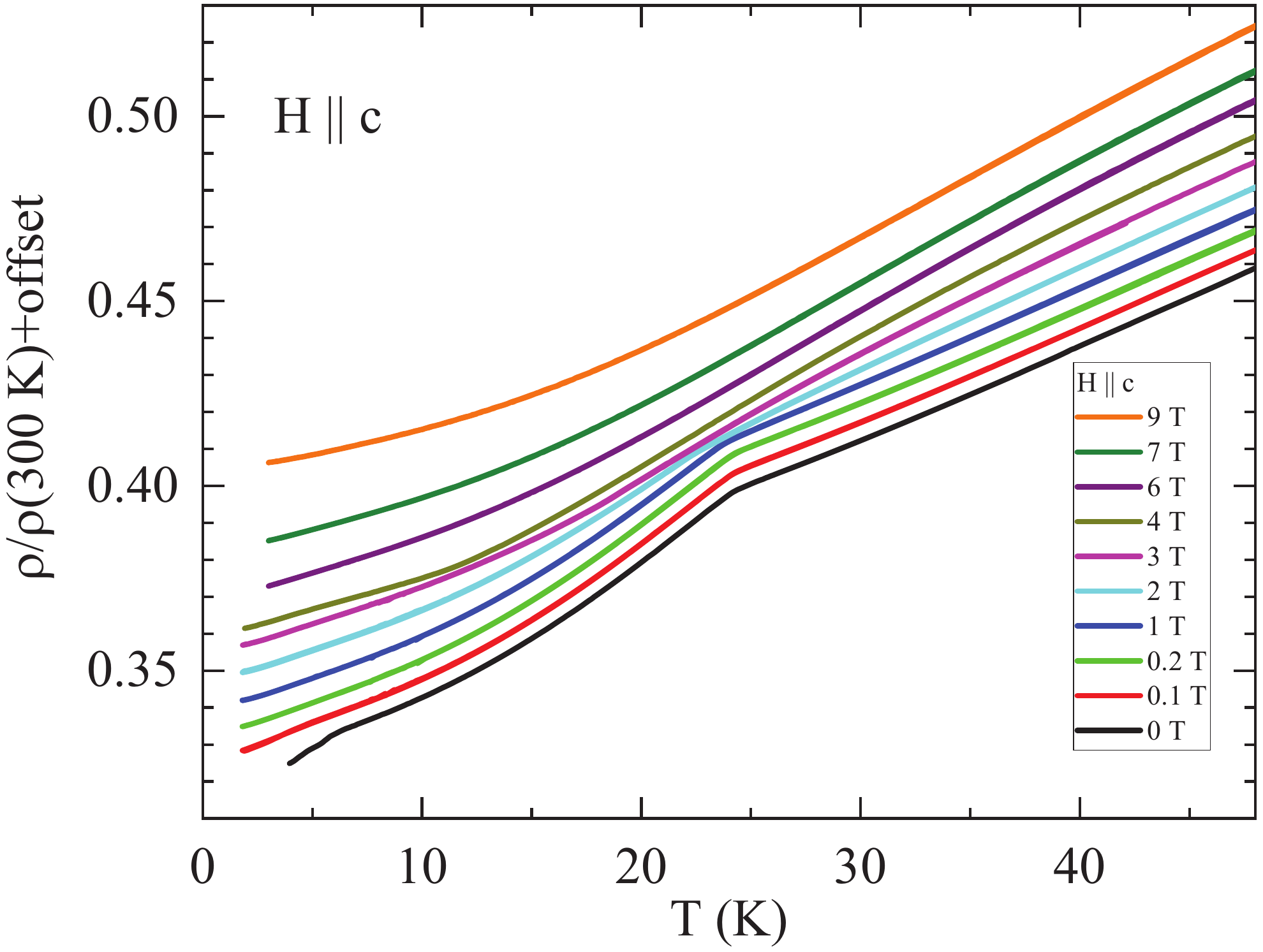}
\caption{Temperature dependence of the normalized resistivity $\rho(T)/\rho(300~\rm K)$of EuSn$_2$As$_2$ measured in magnetic fields $H \parallel c$ as indicated. The data for successive fields are offset from each other for clarity, where the bottom curve is for $H=0$ and the top one is for $H=9$~T\@.}
\label{resHc}
\end{figure}

\begin{figure}
\centering
\includegraphics[width=3.3in]{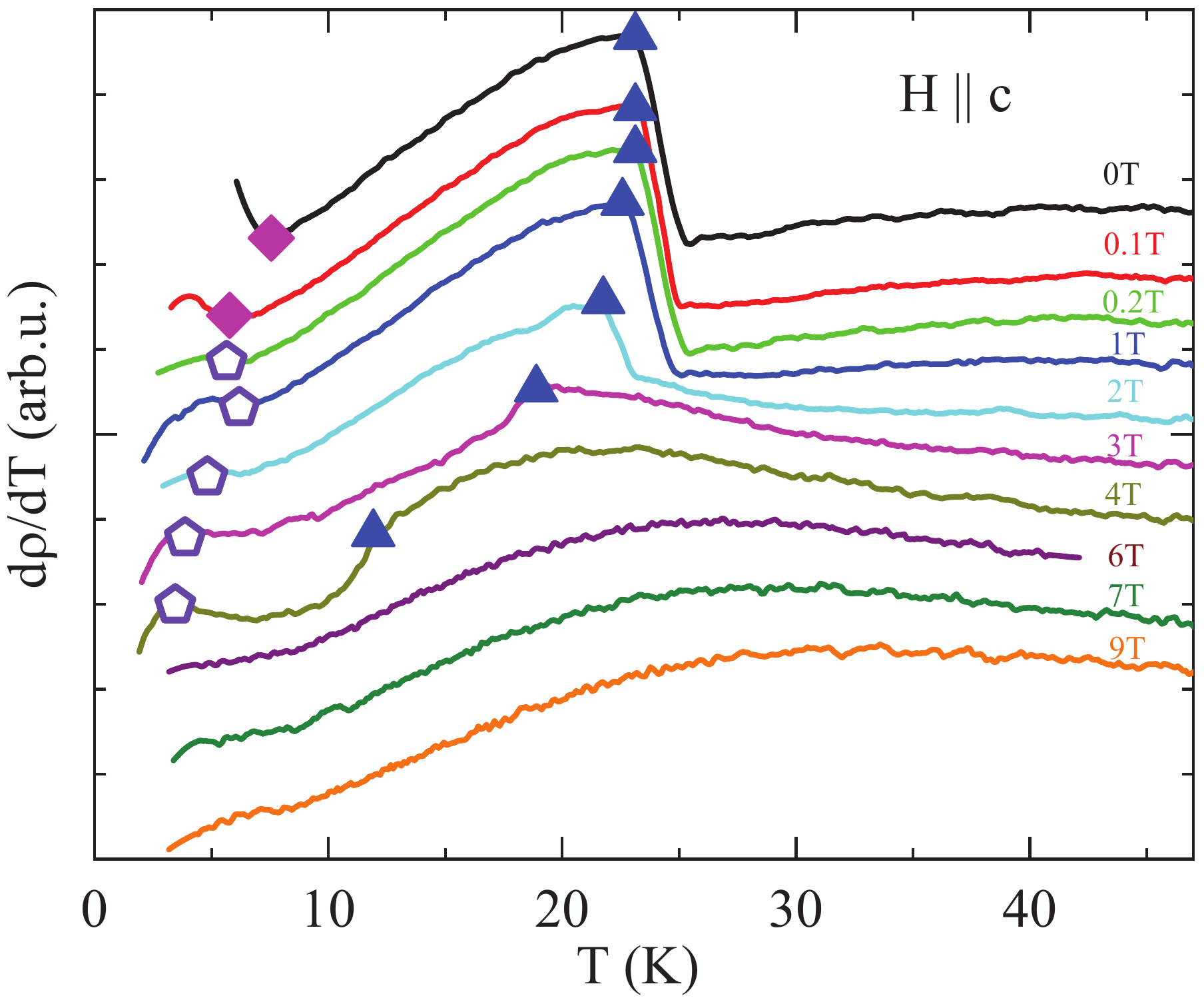}
\caption{Temperature dependence of the normalized resistivity derivative $d[\rho(T)/\rho(300~{\rm K})]/dT$ of EuSn$_2$As$_2$ measured in magnetic fields $H \parallel c$ as indicated.  The filled diamonds and open pentagons at temperatures $T_{\rm A}$ represent potential transitions of unknown origin. The filled blue triangles represent maxima $d[\rho(T)/\rho(300~{\rm K})]/dT$ at $T_{\rm N}(H)$.}
\label{derHc}
\end{figure}

Electrical resistivity $\rho$ measurements were performed on a number of samples from batch C; however, $\rho (T,H)$ of only two samples were studied in detail. They revealed consistent behavior, so we present data for only one of them. The resistivity value at room temperature showed a notable variation, suggesting an important role of cracks~\cite{Tanataranisotropy}. The average $\rho(300$~K) as determined on six samples was 131~$\mu \Omega$\,cm with a standard deviation of 46~$\mu \Omega$\,cm. This value is nearly three times smaller than reported previously~\cite{Arguilla2017_EuSn2As2,Chen2020_EuSn2As2}. Due to this large uncertainty we present normalized resistivity data $\rho/\rho(300 ~\rm K)$\@.

To put the magnitude of $\rho(300$~K) in perspective, a similar value of  150~$\mu \Omega$\,cm was found in the optimally-doped iron-based superconductors~\cite{Tanatarpseudogap}.  Within the error bars, a somewhat smaller but comparable value of $(120\pm30)~\mu \Omega$\,cm was found in SrCo$_2$As$_2$~\cite{Sangeetha2019}. The $\rho(300~{\rm K})$ value of \esa\ and the monotonically-decreasing resistivity on cooling (see the top left inset in Fig.~\ref{resistivity}) suggest a metallic character of the electronic transport with no contribution of Dirac cone carriers. This is natural considering that the Dirac crossing is located 0.1~eV above the Fermi energy~\cite{Li2019_EuSn2As2}. The metallic character of transport is in line with the very weak $T$ dependence of the Hall effect~\cite{Chen2020_EuSn2As2} suggesting hole carriers with concentration $p=3.6\times10^{20}$~cm$^{-3}$ at 2~K and with  less than a 10\% change on increasing $T$ to 200~K\@.
    
The $T$-dependent~$\rho$ of EuSn$_2$As$_2$ was measured in $H=0$ and in constant nonzero~$H$ applied parallel to the conducting $ab$~plane. The resistivity in $H=0$ over the whole $T$ range below room temperature is shown in the top-left inset of Fig.~\ref{resistivity}. The resistivity is close to a $T$-linear dependence at high $T$, as normally expected of metals \cite{BG}, followed by a slight upturn below $T\approx 50$~K and a sharp decrease below $T_{\rm N} \approx 24$~K as observed previously~\cite{Chen2020_EuSn2As2}. This dependence is consistent with a significant contribution of magnetic scattering in the paramagnetic state above $T_{\rm N}$~\cite{MagneticScattering}, revealed also in the negative magnetoresistance at temperatures at least as high as 40~K    associated with spin correlations as also observed in the $C_{\rm p}(T)$ data in Figs.~\ref{Fig_Heat_capacity_MFT}(a) and~\ref{Fig_Heat_capacity_MFT}(b). 

Below $T_{\rm N}$ the resistivity in $H=0$ shows a sharp decrease with a clear anomaly in the $T$-dependent derivative $d \rho/dT$ in Fig.~\ref{derivative}, as expected for magnetic materials~\cite{FisherdrhodT}.  At temperatures below $T_L \sim6$~K, a second anomaly is observed in $\rho(T)$ as a clearly-visible downturn in the bottom-right inset in Fig.~\ref{resistivity} and as a sharp upturn in $d\rho/dT$ in Fig.~\ref{derivative}. These features are rapidly suppressed in magnetic fields as low as 0.05~T\@.

On application of a magnetic field with $H \parallel ab$, the resistivity at temperatures both above and below $T_{\rm N}$ is significantly suppressed. Notably, this suppression persists up to $\approx 40~{\rm K}$, above which the magnetoresistance changes sign from negative to positive,  similar to previous results for $H\parallel c$~\cite{Chen2020_EuSn2As2}.  Positive magnetoresistance is a characteristic of a usual metal, as also reveled in the $T$-inear resistivity above 40~K\@.  The onset temperature of a magnetic effect on $\rho$ on cooling as notably revealed by the sign change of the magnetoresistance correlates well with the temperature of complete entropy release of the Eu$^{2+}$ spins in the heat capacity measurements in Fig.~\ref{Fig_Heat_capacity_MFT}, as also found in CeRhIn$_5$ \cite{PaglioneRh}. 

The negative sign of magnetoresistance below $\approx 40$~K suggests that magnetic scattering is suppressed with magnetic field even in the magnetically-ordered state, revealing significant residual disorder in the magnetic subsystem. This is not expected in simple collinear FM or AFM  materials~\cite{PaglioneQCP,PaglioneRh}, and suggests a more complicated type of magnetic ordering. The zero-field neutron-diffraction and low-field magnetic susceptibility data demonstrate that the AFM ordering is collinear A-type with the moments aligned in the $ab$~plane.  However, in an $ab$-plane field the magnetic structure likely changes due to rotations of the collinear AFM-ordered moments in the three domains to become perpendicular to the field as discussed in Sec.~\ref{Sec:chiM}.

Another feature in the resistivity data supporting an $H_{ab}$-dependent change in the magnetic structure of \esa\ is splitting of the main transition at $T_{\rm N}$\@.  As shown in Fig.~\ref{derivative}, the sharp transtion turns into a broad crossover and a sharp peak develops at about 20~K\@.  The onset of this peak can be noticed already at 1~T\@.  In a field of 3~T the peak transforms into broad double humps, and the curve becomes featureless at $H=4$~T with a broad crossover maximum in the resistivity derivative still centered around 20~K\@. These results correlate with the change in the magnetic structure as the ordered moments, initially perpendicular to the field in $h_{\rm ab} = 0.15$~T, cant towards the applied field, where the critical field for FM alignment of the moments with the field from Fig.~\ref{Fig_M-H_hysteresis} is $\approx 3.7$~T\@.

Resistivity measurements versus $T$ with $H\parallel c$ are presented in Fig.~\ref{resHc}, and Fig.~\ref{derHc} shows the evolution of the \mbox{$T$-dependent} derivative of the resistivity versus~$T$\@. The curves in both figures are offset to avoid overlapping, where the offset is upwards in Fig.~\ref{resHc} but downwards in Fig.~\ref{derHc}. The $\rho(T)$ reveals a complcated transformation from a sharp anomaly at $T_{\rm N}$ in zero field to a broad crossover field at $H=3$~T.  A small feature at $\sim 7$~K transforms into a tiny decrease in the resistivity derivative, suggesting a nearly vertical line in the phase diagram of unclear origin.


\section{\label{Summary} Summary}

\begin{figure*}[ht]
\includegraphics[width = 3.3in]{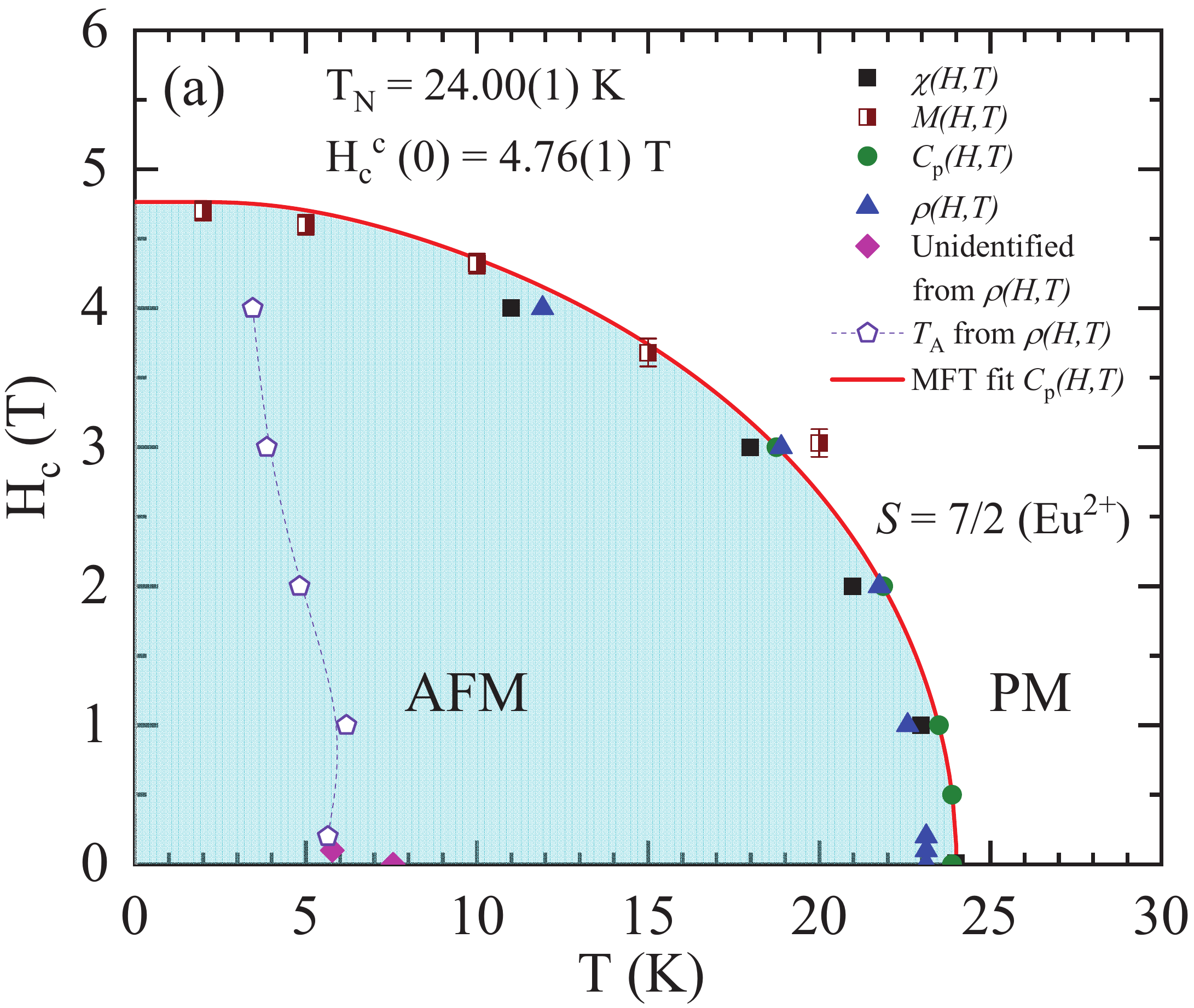}\includegraphics[width = 3.3in]{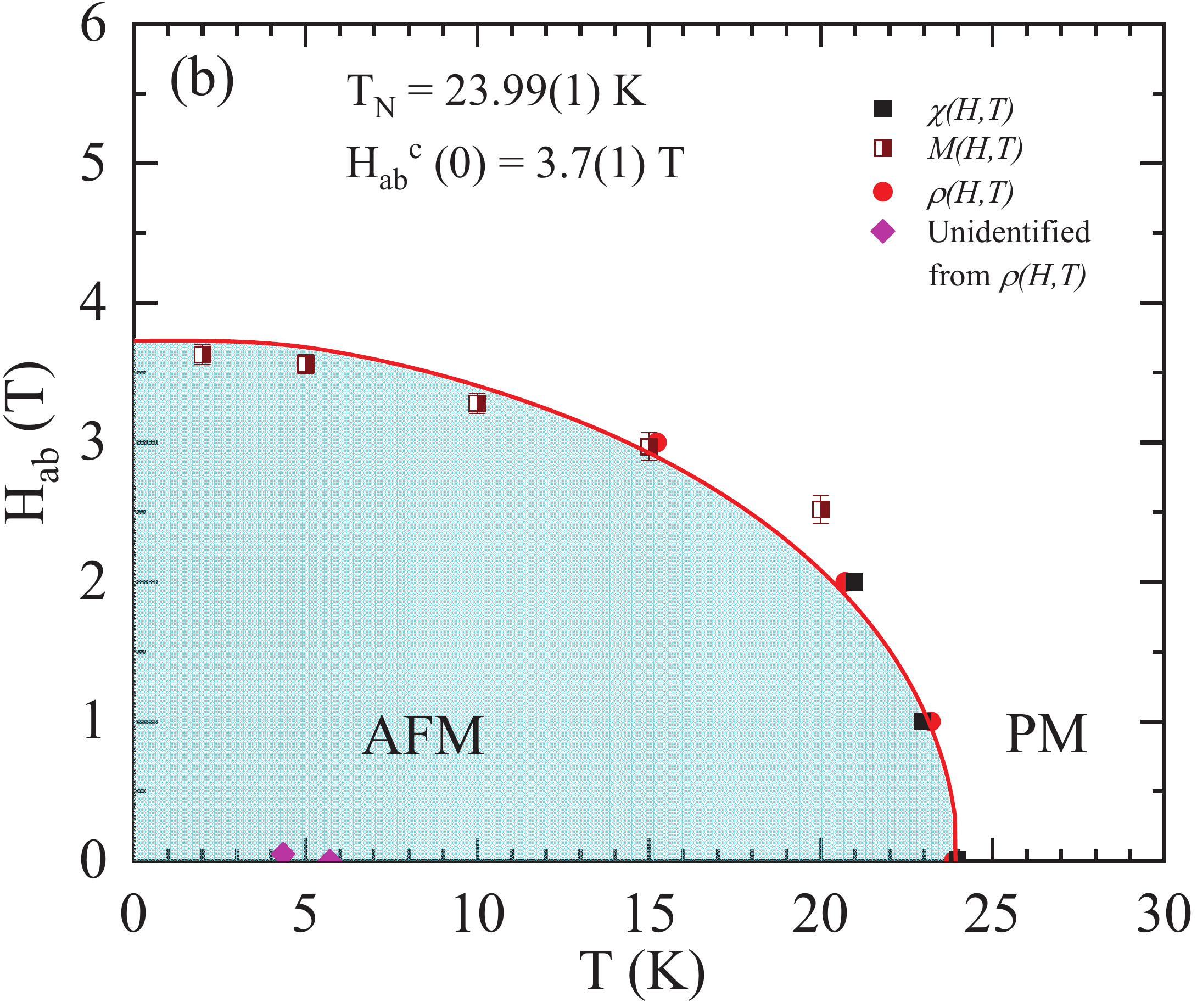}
\caption{(a)~Magnetic $c$-axis field $H_c$ vs $T$ phase diagram of EuSn$_2$As$_2$ constructed from the $C_{\rm p}(H,T)$ (filled green circles), $\chi(H,T)$ (filled black squares), and $\rho(H,T)$ (filled blue triangles) data.  The solid red curve is a fit of the critical field $H_c^{\rm c}(T)$ data obtained from the $C_{\rm p}(T)$ data in Fig.~\ref{Fig_Heat_capacity_Field} by Eq.~(\ref{Eq:phase_diagram_MFT1}) that separates the antiferromagnetic (AFM) phase from the paramagnetic (PM) phase. We suggest that the filled magenta diamonds near zero field near 5 and 6~K identified from the $\rho(H,T)$ measurements are due to lockin transitions of the Eu moments in three AFM domains becoming perpendicular to the applied field.  The unfilled pentagons at temperatures $T_{\rm A}$ are unidentified transitions suggested from the $\rho(H,T)$ measurements in Fig.~\ref{derHc}.    (b)~$ab$-plane magnetic field $H_{ab}$ vs $T$ phase diagram of \esa\ as determined from $\rho(H_{ab},T)$ measurements.  The red filled circles represent $H_{ab}(T_{\rm N})$ obtained from Fig.~\ref{derivative}.  The violet diamonds near $H_{ab}=0$ correspond to a feature in the $T$-dependent resistivity derivative which we suggest reflects a lock-in transition at which the Eu moments in the $ab$~plane become perpendicular to the applied field. The solid magenta diamonds at 3--7~K  correspond to a small anomaly observed in $\rho(T)$ of unknown origin.  The critical fields $H^{\rm c}$ vs $T$ in (a) and~(b) (half-filled boxes) from Table~\ref{Tab.Criticalfield} were determined as the intersection of the extrapolated lines from larger $H$ and smaller $H$ with respect to the approximate saturation field. }
\label{Fig_Phase_diagram}
\end{figure*}

We investigated the ground state spin texture and the $H$-$T$ phase diagrams of the AFM topological insulator material \esa\ which has recently been found to host Dirac surface states. The compound undergoes AFM ordering below the N\'eel temperature $T_{\rm N} = 24$~K as identified from neutron diffraction measurements.  The magnetic susceptibility $\chi(T)$ measurements find the same $T_{\rm N}$ with an additional low-$T$ upturn below $\sim 10$~K that is evidently due to magnetic impurities or defects in the crystals.  By measuring the $\chi(T)$ on three \esa\ crystals grown differently, we showed that the low-$T$ upturn in the $\chi(T)$ data is crystal-growth dependent. We subsequently examined the physical properties of crystals with the lowest concentration of magnetic impurities/defects (Crystals~C).

By analyzing $\chi(T)$ data measured at a low magnetic field of 0.01~T on the basis of molecular field theory together with zero-field neutron-diffraction measurements, we established that the AFM structure of the Eu$^{2+}$ spins $S = 7/2$ below $T_{\rm N}$ is collinear A-type. Here the Eu spins within an $ab$~plane are aligned ferromagnetically within the plane and aligned antiferromagnetically between adjacent $ab$-plane layers along the $c$~axis.  

The in-plane magnetic susceptibility $\chi_{ab} = M_{ab}/H$ at 2~K was found to be sensitive to the applied magnetic field in the low-$H$ region, while the out-of-plane $\chi_{c}=M_c/H$ is approximately independent of $H$ up to the $c$-axis critical field $H_c^{\rm c}$.  The former measurement exhibits a  smooth maximum in $dM_{ab}/dH$ at $H \approx 0.12$~T which we interpret as the in-plane Eu moments in different threefold domains locking in to become perpendicular to the applied field. At higher fields, the $ab$-plane magnetization increases due to the usual canting of the moments toward the field direction.

A $\lambda$ anomaly is observed in the heat capacity at $T_{\rm N}=24$~K that is gradually suppressed with increasing $H$\@. The full magnetic entropy associated with the Eu$^{2+}$ spins is released by about 60~K, which is more than twice $T_{\rm N}$, implying the presence of strong short-range dynamic AFM spin correlations above $T_{\rm N}$\@. 

The $H_c$-$T$ magnetic phase diagram  shown  in Fig.~\ref{Fig_Phase_diagram}(a) was constructed from the combined results of $T_{\rm N}(H)$ data obtained from $c$-axis $\chi(H,T)$ (black solid squares), heat capacity $C_{\rm p}(T)$ (green solid circles), and electrical resistivity$\rho(H,T)$ (blue up-triangles) measurements. The phase boundary between the AFM and PM states is consistent with the molecular-field-theory prediction for spin~$S=7/2$.  According to MFT, at a fixed $T$ the critical field $H_{c}^{\rm c}$ responsible for an AFM to PM phase transition in a Heisenberg antiferromagnet with $ab$-plane moment alignment is given by~\cite{Johnston2015}
\bea
\label{Eq:phase_diagram_MFT1}
H_{c}^{\rm c}(t) = H_{c}^{\rm c}(0)\bar{\mu}_0(t)
\eea
where the symbols and their evaluation are described in Eqs.~(\ref{Eqs:Chixy}).  The red solid curve in Fig.~\ref{Fig_Phase_diagram}(a) shows the fit by Eq.~(\ref{Eq:phase_diagram_MFT1}) of the $H_{c}^{\rm c}(T)$ vs $T_{\rm N}$ data obtained from $C_{\rm p}(H,T)$ measurements. The best fit is obtained with $T_{\rm N} = 24.00(1)$~K and \mbox{$H_{c}^{\rm c}(T = 0) = 4.76(1)~{\rm T}$} as listed in the figure. 

The magnetic phase diagram in Fig.~\ref{Fig_Phase_diagram}(b) for fields in the $ab$-plane was constructed from $ab$-plane $\rho(H_{ab},T)$ measurements in Fig.~\ref{derivative}.   The inset shows a features at $T = 6$ and 5~K in $H=0.05$ and 0.1~T, respectively, that are suppressed in higher fields and transform into a small resistivity decrease feature.  These features are suggested to be associated with the lock-in transition of the Eu moments becoming perpendicular to $H_{ab}$. These features complement the collinear AFM ordering found from the zero-field neutron-diffraction measurements. 


\acknowledgments

This research was supported by the U.S. Department of Energy, Office of Basic Energy Sciences, Division of Materials Sciences and Engineering. Ames Laboratory is operated for the U.S. Department of Energy by Iowa State University under Contract No.~DE-AC02-07CH11358.



\begin{thebibliography}{99}

\bibitem{Fertig1977_IC} W. A. Fertig, D. C. Johnston, L. E. DeLong, R. W. McCallum, M. B. Maple, and B. T. Matthias, Destruction of Superconductivity at the Onset of Long-Range Magnetic Order in the Compound ErRh$_4$B$_4$, Phys. Rev. Lett. {\bf 38}, 987 (1977).

\bibitem{Nagarajan1994_IC} R. Nagarajan, C, Mazumdar, Z. Hossain, S. K. Dhar, K. V. Gopalakrishnan, L. C. Gupta, C. Godart, B. D. Padalia, and R. Vijayaraghavan, Bulk superconductivity at an elevated temperature ($T_c\cong 12$ K) in a nickel containing alloy system Y-Ni-B-C, Phys. Rev. Lett. {\bf 72}, 274 (1994); {\bf 73}, 211(E) (1994).

\bibitem{Ghosh1995_IC} K. Ghosh, S. Ramakrishnan, S. K. Dhar, S. K. Malik, Girish Chandra, V. K. Pecharsky, K. A. Gschneidner, Jr., Z. Hu, and W. B. Yelon, Crystal structures and low-temperature behaviors of the heavy-fermion compounds CeRuGe$_3$ and $\rm{Ce_3Ru_4Ge_{13}}$ containing both trivalent and tetravalent cerium, Phys. Rev. B {\bf 52}, 7267 (1995).

\bibitem{Curro2000_IC} N. J. Curro, P. C. Hammel, P. G. Pagliuso, J. L. Sarrao, J. D. Thompson, and Z. Fisk, Evidence for spiral magnetic order in the heavy fermion material CeRhIn$_5$, Phys. Rev. B {\bf 62}, R6100 (2000).

\bibitem{Hundley2001_IC} M. F. Hundley, J. L. Sarrao, J. D. Thompson, R. Movshovich, M. Jaime, C. Petrovic, and Z. Fisk, Unusual Kondo behavior in the indium-rich heavy-fermion antiferromagnet $\rm{Ce_3Pt_4In_{13}}$, Phys. Rev. B {\bf 65}, 024401 (2001).

\bibitem{Sakai2011_IC} A. Sakai and S. Nakatsuji, Strong valence fluctuation effects in Sm$Tr_2$Al$_{20}$  ($Tr$ = Ti, V, Cr), Phys. Rev. B {\bf 84}, 201106(R) (2011).

\bibitem{Yamaoka2014_IC} H. Yamaoka, Y. Ikeda, I. Jarrige, N. Tsujii, Y. Zekko, Y. Yamamoto, J. Mizuki, J.-F. Lin, N. Hiraoka, H. Ishii, K.-D. Tsuei, T. C. Kobayashi, F. Honda, and Y. Onuki, Role of Valence Fluctuations in the Superconductivity of Ce122 Compounds, Phys. Rev. Lett. {\bf 113}, 086403 (2014).

\bibitem{Pecharsky1997_IC} V. K. Pecharsky and K. A. Gschneidner, Jr., Giant Magnetocaloric Effect in $\rm{Gd_5(Si_2Ge_2)}$, Phys. Rev. Lett. {\bf 78}, 4494 (1997).

\bibitem{Pakhira2016_IC} S. Pakhira, C. Mazumdar, R. Ranganathan, S. Giri, and M. Avdeev, Large magnetic cooling power involving frustrated antiferromagnetic spin-glass state in $R_2$NiSi$_3$ ($R$ = Gd, Er), Phys. Rev. B {\bf 94}, 104414 (2016).

\bibitem{Pakhira2017_IC} S. Pakhira, C. Mazumdar, R. Ranganathan, and M. Avdeev, Magnetic frustration induced large magnetocaloric efect in the
absence of long range magnetic order, Sci. Rep. {\bf 7}, 7367 (2017).

\bibitem{Buschow1969_IC} K. H. J. Buschow and H. J. van Daal, Evidence for the Presence of the Kondo effect in the Compound CeAl$_2$, Phys. Rev. Lett. {\bf 23}, 408 (1969).

\bibitem{Gignoux1984_IC} D. Gignoux and J. C. Gomez-Sal, Competition between the Kondo effect and exchange interactions in the CeNi$_x$Pt$_{1 - x}$ compounds, Phys. Rev. B {\bf 30}, 3967 (1984).

\bibitem{Ishida2002_IC} K. Ishida, K. Okamoto, Y. Kawasaki, Y. Kitaoka, O. Trovarelli, C. Geibel, and F. Steglich, $\rm{YbRh_2Si_2}$: Spin Fluctuations in the Vicinity of a Quantum Critical Point at Low Magnetic Field, Phys. Rev. Lett. {\bf 89}, 107202 (2002).

\bibitem{Arndt2011_IC} J. Arndt, O. Stockert, K. Schmalzl, E. Faulhaber, H. S. Jeevan, C. Geibel, W. Schmidt, M. Loewenhaupt, and F. Steglich, Spin Fluctuations in Normal State $\rm{CeCu_2Si_2}$ on Approaching the Quantum Critical Point, Phys. Rev. Lett. {\bf 106}, 246401 (2011).

\bibitem{Hirschberger2016_GdPtBi} M. Hirschberger, S. Kushwaha, Z. Wang, Q. Gibson, S. Liang, C. A. Belvin, B. A. Bernevig, R. J. Cava, and N. P. Ong, The chiral anomaly and thermopower of Weyl fermions in the half-Heusler GdPtBi, Nat. Mater. {\bf 15}, 1161 (2016).

\bibitem{Shekhar2018_RPtBi} C. Shekhar, N. Kumar, V. Grinenko, S. Singh, R. Sarkar, H. Luetkens, Shu-Chun Wu, Y. Zhang, A. C. Komarek, E.Kampert, Y. Skourski, J. Wosnitza, W. Schnelle, A. McCollam, U. Zeitler, J. K\"{u}bler, B. Yan, H.-H. Klauss, S. S. P. Parkin, and C. Felser, Anomalous Hall effect in Weyl semimetal half-Heusler compounds RPtBi (R = Gd and Nd), PNAS {\bf 115}, 9140 (2018).

\bibitem{Wang2016_YbMnBi2} A. Wang, I. Zaliznyak, W. Ren, L. Wu, D. Graf, V. O. Garlea, J. B. Warren, E. Bozin, Y. Zhu, and C. Petrovic, Magnetotransport study of Dirac fermions in YbMnBi$_2$ antiferromagnet, Phys. Rev. B {\bf 94}, 165161 (2016).

\bibitem{May2014_EuMnBi2} A. F. May, M. A. McGuire, and B. C. Sales, Effect of Eu magnetism on the electronic properties of the candidate Dirac material EuMnBi$_2$, Phys. Rev. B {\bf 90}, 075109 (2014).

\bibitem{Masuda2016_EuMnBi2} H. Masuda, H. Sakai, M. Tokunaga, Y. Yamasaki, A. Miyake, J. Shiogai, S. Nakamura, S. Awaji, A. Tsukazaki, H. Nakao, Y. Murakami, T. Arima, Y. Tokura, and S. Ishiwata, Quantum Hall effect in a bulk antiferromagnet EuMnBi$_2$ with magnetically confined two-dimensional Dirac fermions, Sci. Adv. {\bf 2}, e1501117 (2016).

\bibitem{Soh2019_EuMnSb2} J.-R. Soh, P. Manuel, N. M. B. Schr\"{o}ter, C. J. Yi, F. Orlandi, Y. G. Shi, D. Prabhakaran, and A. T. Boothroyd, Magnetic and electronic structure of Dirac semimetal candidate EuMnSb$_2$, Phys. Rev. B {\bf 100}, 174406 (2019).

\bibitem{Schellenberg2011_EuCd2As2} I. Schellenberg, U. Pfannenschmidt, M. Eul, C. Schwickert, and R. P\"{o}ttgen, A $^{121}$Sb and $^{151}$Eu M\"{o}ssbauer Spectroscopic Investigation of EuCd$_2X_2$ ($X$ = P, As, Sb) and YbCd$_2$Sb$_2$, Z. Anorg. Allg. Chem. {\bf 637}, 1863 (2011).

\bibitem{Jo2020_EuCd2As2} N. H. Jo, B. Kuthanazhi, Y. Wu, E. Timmons, T.-H. Kim, L. Zhou, L.-L. Wang, B. G. Ueland, A. Palasyuk, D. H. Ryan, R. J. McQueeney, K. Lee, B. Schrunk, A. A. Burkov, R. Prozorov, S. L. Bud'ko, A. Kaminski, and P. C. Canfield, Manipulating of magnetism in the topological semimetal EuCd$_2$As$_2$, Phys. Rev. B {\bf 101},  140402(R) (2020).

\bibitem{Xu2019_EuIn2As2} Y. Xu, Z. Song, Z. Wang, H. Weng, and X. Dai, Higher-Order Topology of the Axion Insulator EuIn$_2$As$_2$, Phys. Rev. Lett. {\bf 122}, 256402 (2019).

\bibitem{Pakhira2020_EuMg2Bi2} S. Pakhira, M. A. Tanatar, and D. C. Johnston, Magnetic, thermal, and electronic-transport properties of \emb\ single crystals, Phys. Rev. B {\bf 101}, 214407 (2020).

\bibitem{Pakhira2021_EuMg2Bi2} S. Pakhira, T. Heitmann, S. X. M. Riberolles, B. G. Ueland, R. J. McQueeney, D. C. Johnston, and D. Vaknin, Zero-field magnetic ground state of \emb , Phys. Rev. B {\bf 103}, 024408 (2021).

\bibitem{Hasan2010_TI} M. Z. Hasan and C. L. Kane, \textit{Colloquium}: Topological insulators, Rev. Mod. Phys. {\bf 82}, 3045 (2010).

\bibitem{Fu2007_TI} L. Fu and C. L. Kane, Topological insulators with inversion symmetry, Rev. Mod. Phys. {\bf 76}, 045302 (2007).

\bibitem{Tokura2019_TI} Y. Tokura, K. Yasuda, and A. Tsukazaki, Magnetic topological insulators, Nat. Rev. Phys. {\bf 1}, 126 (2019).

\bibitem{Qi2006_QAHE} X.-L. Qi, Y.-S. Wu, and S.-C. Zhang, Topological quantization of the spin Hall effect in two-dimensional paramagnetic semiconductors, Phys. Rev. B {\bf 74}, 085308 (2006).

\bibitem{Liu2008_QAHE} C.-X. Liu, X.-L. Qi, X. Dai, Z. Fang, and S.-C. Zhang, Quantum Anomalous Hall Effect in
Hg$_{1-y}$Mn$_y$Te Quantum Wells, Phys. Rev. Lett. {\bf 101}, 146802 (2008).

\bibitem{Yu2010_QAHE} R. Yu, W. Zhang, H.-J. Zhang, S.-C. Zhang, X. Dai, and Z. Fang, Quantized Anomalous Hall Effect in Magnetic Topological Insulators, Science {\bf 329}, 61 (2010).

\bibitem{Essin2009_AED} A. M. Essin, J. E. Moore, and D. Vanderbilt, Magnetoelectric Polarizability and Axion Electrodynamics in Crystalline Insulators, Phys. Rev. Lett. {\bf 102}, 146805 (2009).

\bibitem{Li2010_AED} R. Li, J. Wang, X.-L. Qi and S.-C. Zhang, Dynamical axion field in topological magnetic insulators, Nat. Phys. {\bf 6}, 284 (2010).

\bibitem{Fu2008_Majorana} L. Fu and C. L. Kane, Superconducting Proximity Effect and Majorana Fermions at the Surface of a Topological Insulator, Phys. Rev. Lett. {\bf 100}, 096407 (2008).

\bibitem{Akhmerov2009_Majorana} A. R. Akhmerov, J. Nilsson, and C. W. J. Beenakker, Electrically Detected Interferometry of Majorana Fermions in a Topological Insulator, Phys. Rev. Lett. {\bf 102}, 216404 (2009).

\bibitem{Cook2011_Majorana} A. Cook and M. Franz, Majorana fermions in a topological-insulator nanowire proximity-coupled to an ${\bf\it s}$-wave superconductor, Phys. Rev. B {\bf 84}, 201105(R) (2011).

\bibitem{Li2019_EuSn2As2} H. Li, S.-Y. Gao, S.-F. Duan, Y.-F. Xu, K.-J. Zhu, S.-J. Tian, J.-C. Gao, W.-H. Fan, Z.-C. Rao, J.-R. Hugang, J.-J. Li, D.-Y. Yan, Z.-T. Liu, W.-L. Liu, Y.-B. Huang, Y.-L. Li, Y. Liu, G.-B. Zhang, P. Zhang, T. Kondo, S. Shin, H.-C. Lei, Y.-G. Shi, W.-T. Zhang, H.-M. Weng, T. Qian, and H. Ding, Dirac Surface States in Intrinsic Magnetic Topological Insulators EuSn$_2$As$_2$ and MnBi$_{2n}$Te$_{3n+1}$, Phys. Rev. X {\bf 9}, 041039 (2019).

\bibitem{Chen2020_EuSn2As2} H.-C. Chen, Z.-F. Lou, Y.-X. Zhou, Q. Chen, B.-J. Xu, S.-J. Chen, J.-H. Du, J.-H. Yang, H.-D. Wang, and M.-H. Fang, Negative Magnetoresistance in Antiferromagnetic Topological Insulator EuSn$_2$As$_2$, Chinese Phys. Lett. {\bf 37}, 047201 (2020).

\bibitem{Arguilla2017_EuSn2As2} M. Q. Arguilla,   N. D. Cultrara,   Z. J. Baum,   S. Jiang,   R. D. Ross  and  J. E. Goldberger, EuSn$_2$As$_2$: An Exfoliatable Magnetic Layered Zintl-Klemm Phase, Inorg. Chem. Front. {\bf 4}, 378 (2017).

\bibitem{Carvajal1993} J. Rodr\'{\i}guez-Carvajal, Recent advances in magnetic structure determination by neutron powder diffraction, Physica B {\bf 192}, 55 (1993).

\bibitem{Tanataranisotropy} M. A. Tanatar, N. Ni, C. Martin, R. T. Gordon, H. Kim, V. G. Kogan, G. D. Samolyuk, S. L. Bud'ko, P. C. Canfield, and R. Prozorov,
Anisotropy of the iron pnictide superconductor Ba(Fe$_{1-x}$Co$_x$)$_2$As$_2$, ($x=0.074$,
$T_{\rm c} = 23$~K), 
Phys. Rev. B {\bf 79}, 094507 (2009).

\bibitem{SUST} M. A. Tanatar, N. Ni, S. L. Bud'ko, P. C. Canfield, and R. Prozorov, Field-dependent transport critical current in single crystals of Ba(Fe$_{1-x}$TM$_x$)$_2$As$_2$ (TM = Co, Ni) superconductors, Supercond. Sci. Technol. {\bf 23}, 054002 (2010).

\bibitem{Momma2011} K. Momma and F. Izumi, \textit{VESTA 3} for three-dimensional visualization of crystal, volumetric and morphology data, J. Appl. Cryst. {\bf 44}, 1272 (2011).

\bibitem{Asbrand1995} M. Asbrand, B. Eisenmann, and J. Klein, Arsenidostannates with [SnAs] Nets Isostructural to Grey Arsenic:  Synthesis and Crystal Structure of Na[Sn$_2$As$_2$], Na$_{0.3}$Ca$_{0.7}$[Sn$_2$As$_2$], Na$_{0.4}$Sr$_{0.6}$[Sn$_2$As$_2$], Na$_{0.6}$Ba$_{0.4}$[Sn$_2$As$_2$], and K$_{0.3}$Sr$_{0.7}$[Sn$_2$As$_2$],  Z. anorg. allg. Chem. {\bf 621}, 576 (1995).

\bibitem{Johnston2012} D. C. Johnston, Magnetic Susceptibility of Collinear and Noncollinear Heisenberg Antiferromagnets, Phys. Rev. Lett. {\bf 109}, 077201 (2012).

\bibitem{Johnston2015} D. C. Johnston, Unified molecular field theory for collinear and noncollinear Heisenberg antiferromagnets, Phys. Rev. B {\bf 91}, 064427 (2015).

\bibitem{Pakhira2020} S. Pakhira, M. A. Tanatar, and D. C. Johnston, Magnetic, thermal, and electronic-transport properties of ${\rm EuMg_2Bi_2}$ single crystals, Phys. Rev. B {\bf 101}, 214407 (2020).

\bibitem{Pakhira2021} S. Pakhira, T. Heitmann, S. X. M. Riberolles, B. G. Ueland, R. J. McQueeney, D. C. Johnston, and D. Vaknin, Zero-field magnetic ground state of ${\rm EuMg_2Bi_2}$, Phys. Rev. B {\bf 103}, 024408 (2021).

\bibitem{Johnston2016} D. C. Johnston, Magnetic dipole interactions in crystals, Phys. Rev. B {\bf 93}, 014421 (2016).

\bibitem{Johnston2017} D. C. Johnston, Magnetic structure and magnetization of helical antiferromagnets in high magnetic fields perpendicular to the helix axis at zero temperature, Phys. Rev. B {\bf 96}, 104405 (2017); {\bf 98}, 099903(E) (2018).

\bibitem{Goetsch_2012} R. J. Goetsch, V. K. Anand, A. Pandey, and D. C. Johnston, Structural, thermal, magnetic, and electronic transport properties
of the LaNi$_2$(Ge$_{1-x}$P$_x$)$_2$ system, Phys. Rev. B {\bf 85}, 054517 (2012).

\bibitem{Sangeetha_EuCo2As2_2018} N. S. Sangeetha, V. K. Anand, E. Cuervo-Reyes, V. Smetana, A.-V. Mudring, and D. C. Johnston, Enhanced moments of Eu in single crystals of the metallic helical antiferromagnet EuCo$_{2-y}$As$_2$, Phys. Rev. B {\bf 97}, 144403 (2018).

\bibitem{Sangeetha_EuCo2P2_2016} N. S. Sangeetha, E. Cuervo-Reyes, A. Pandey, and D. C. Johnston, EuCo$_2$P$_2$: A model molecular-field helical Heisenberg antiferromagnet, Phys. Rev. B {\bf 94}, 014422 (2016).

\bibitem{Sangeetha_EuNi2As2_2019} N. S. Sangeetha, V. Smetana, A.-V. Mudring, and D. C. Johnston, Helical antiferromagnetic ordering in EuNi$_{1.95}$As$_2$ single crystals, Phys. Rev. B {\bf 100}, 094438 (2019).

\bibitem{Anand2015} V. K. Anand and D. C. Johnston, Antiferromagnetism in EuCu$_2$As$_2$ and EuCu$_{1.82}$Sb$_2$ single crystals, Phys. Rev. B {\bf 91}, 184403 (2015).

\bibitem{Anand2014} V. K. Anand and D. C. Johnston, Physical properties of ${\rm EuPd_2As_2}$ single crystals, J. Phys.: Condens. Matter {\bf 26}, 282002 (2014).

\bibitem{Singh2009} Y. Singh, Y. Lee, B. N. Harmon, and D. C. Johnston, Unusual magnetic, thermal, and transport behavior of single-crystalline ${\rm EuRh_2As_2}$, Phys. Rev. B {\bf 79}, 220401(R) (2009).

\bibitem{Pandey2017} A. Pandey, C. Mazumdar, R Ranganathan, and D. C. Johnston, Multiple crossovers between positive and negative magnetoresistance versus field due to fragile spin structure in metallic ${\rm GdPd_3}$, Sci. Rep. {\bf 7}, 42789 (2017).

\bibitem{Johnston2021} D. C. Johnston, Molecular-field-theory fits to magnetic susceptibilities of antiferromagnetic ${\rm GdCu_2Si_2}$, CuO, ${\rm LiCrO_2}$, and $\alpha$-${\rm CaCr_2O4}$ single crystals below their N\'eel temperatures, J. Magn. Magn. Mater. {\bf 535}, 168062 (2021).

\bibitem{Tanatarpseudogap} 
M. A. Tanatar, N. Ni, A. Thaler, S. L. Bud'ko, P. C. Canfield, and R. Prozorov, Pseudogap and its critical point in the heavily doped Ba(Fe$_{1-x}$Co$_x$)$_2$As$_2$ from $c$-axis resistivity measurement, Phys. Rev. B {\bf 82}, 134528 (2010). 

\bibitem{Sangeetha2019} N. S. Sangeetha, L.-L. Wang, A. V. Smirnov, V. Smetana, A.-V. Mudring, D. D. Johnson, M. A. Tanatar, R. Prozorov, and D. C. Johnston, Non-Fermi-liquid types of behavior associated with a magnetic quantum critical point in Sr(Co$_{1-x}$Ni$_x)_2$ single crystals, Phys. Rev. B {\bf 100}, 094447 (2019).

\bibitem{BG} See, for example, J. Bardeen,  Electrical conductivity of metals, J. Appl. Phys. {\bf 11}, 88, (1940).

\bibitem{MagneticScattering}
J. Paglione, M. A. Tanatar, D. G. Hawthorn, R. W. Hill, F. Ronning, M. Sutherland, L. Taillefer, C. Petrovic, and P. C. Canfield, 
Heat Transport as a Probe of Electron Scattering by Spin Fluctuations: The Case of Antiferromagnetic  CeRhIn$_5$, Phys. Rev. Lett. {\bf 94}, 216602 (2005).

\bibitem{FisherdrhodT}
M. E. Fisher and J. S. Langer, Resistive Anomalies at Magnetic Critical Points, Phys. Rev. Lett. {\bf 20}, 665 (1968).

\bibitem{PaglioneRh}
J. Paglione, M. A. Tanatar, D. G. Hawthorn, R. W. Hill, F. Ronning, M. Sutherland, L. Taillefer, C. Petrovic, and P. C. Canfield, Heat Transport as a Probe of Electron Scattering by Spin Fluctuations: The Case of Antiferromagnetic CeRhIn$_5$, Phys. Rev. Lett. {\bf 94}, 216602 (2005).

\bibitem{PaglioneQCP}
J. Paglione, M. A. Tanatar, D. G. Hawthorn, E. Boaknin, R. W. Hill, F. Ronning, M. Sutherland, L. Taillefer, C. Petrovic, and P. C. Canfield, Field-Induced Quantum Critical Point in CeCoIn$_5$, Phys. Rev. Lett. {\bf 91}, 246405 (2003).

\end{thebibliography}
\end{document}